\documentclass[utf8]{arxiv} %

\usepackage{url,hyperref,lineno,microtype,subcaption}
\usepackage{booktabs}

\def\firstAuthorLast{Lebert {et~al.}} %
\def\Authors{Jan Lebert\,$^{1}$, Namita Ravi\,$^{1,2}$, Flavio H. Fenton\,$^{3}$ and Jan Christoph\,$^{1*}$}
\def\Address{
$^{1}$Cardiovascular Research Institute, University of California, San Francisco, USA\\
$^{2}$Yale School of Medicine, Yale University, New Haven, USA\\
$^{3}$School of Physics, Georgia Institute of Technology, Atlanta, USA
}

\def\corrAuthor{Jan Christoph}

\def\corrEmail{jan.christoph@ucsf.edu}

\newcommand{\stimes}{{\times}}

\begin{document}
\onecolumn
\firstpage{1}

\title[Cardiac Phase Mapping using Deep Neural Networks]{Rotor Localization and Phase Mapping of Cardiac Excitation Waves using Deep Neural Networks}

\author[\firstAuthorLast ]{\Authors} %
\address{} %
\correspondance{} %

\extraAuth{}%

\maketitle

\begin{abstract}

The analysis of electrical impulse phenomena in cardiac muscle tissue is important for the diagnosis of heart rhythm disorders and other cardiac pathophysiology.
Cardiac mapping techniques acquire local temporal measurements and combine them to visualize the spread of electrophysiological wave phenomena across the heart surface.
However, low spatial resolution, sparse measurement locations, noise and other artifacts make it challenging to accurately visualize spatio-temporal activity.
For instance, electro-anatomical catheter mapping is severely limited by the sparsity of the measurements, and optical mapping is prone to noise and motion artifacts.
In the past, several approaches have been proposed to obtain more reliable maps from noisy or sparse mapping data.
Here, we demonstrate that deep learning can be used to compute phase maps and detect phase singularities {in optical mapping videos of ventricular fibrillation, as well as in very noisy, low-resolution and extremely sparse simulated data of reentrant wave chaos mimicking catheter mapping data.}
The self-supervised deep learning approach is fundamentally different from classical phase mapping techniques. Rather than encoding a phase signal from time-series data, a deep neural network instead learns to directly associate phase maps and the positions of phase singularities with short spatio-temporal sequences of electrical data. 
{We tested several neural network architectures, based on a convolutional neural network (CNN) with an encoding and decoding structure, to predict phase maps or rotor core positions either directly or indirectly via the prediction of phase maps and a subsequent classical calculation of phase singularities.
Predictions can be performed across different data, with models being trained on one species and then successfully applied to another, or being trained solely on simulated data and then applied to experimental data.}
Neural networks are a promising alternative to conventional phase mapping and rotor core localization methods.
Future uses may include the analysis of optical mapping studies in basic cardiovascular research, as well as the mapping of atrial fibrillation in the clinical setting.

\end{abstract}

\textbf{Keywords:} Spiral waves, complex systems, chaos, heart rhythm disorders, phase singularities, rotor mapping, neural networks, artificial intelligence in cardiology

\newpage

\section{Introduction}
Cardiac muscle cells constantly oscillate between an 'excited' and a 'resting' electrical state, allowing us to assign a phase $\phi$ to the state of each cell during this cycle.
Cardiac mapping techniques, such as catheter electrode mapping or voltage-sensitive optical mapping, measure the spread of electrical impulses across the heart surface and visualize the spatio-temporal evolution of electrical activity. 
These visualizations are frequently depicted as phase maps $\phi (\vec{x},t)$, which uniquely represent the time course of the action potential in each location of the tissue and express the synchronicity of the activation in both space and time. 
Phase maps are particularly suited to characterize the spatio-temporal disorganization of the electrical wave dynamics underlying cardiac fibrillation \citep{Winfree1989, Gray1998, Witkowski1998, Nash2006, Umapathy2010, Christoph2018}. 
During fibrillation, the heart's electrophysiology degenerates into a dynamic state driven by chaotic wave phenomena, which propagate rapidly through the heart muscle and cause irregular, asynchronous contractions.
These inherently three-dimensional wave phenomena can be observed on the heart's surface using optical mapping, where they often take the shape of rapidly rotating spiral vortex waves or 'rotors'.
Phase maps depict these rotors as pinwheel patterns, with each pinwheel consisting of lines of equal phase that merge at the rotational center of the vortex wave. 
The topological defect at the vortex's core is referred to as a phase singularity.
During ventricular fibrillation, phase singularities move across the heart surface, interact with each other, and undergo pairwise creation and annihilation.
Phase singularities provide a means to automatically localize and track reentrant vortex waves through the heart muscle. They can be used to track wavebreaks \citep{Liu2003, Zaitsev2003}, or interactions of vortex cores with the underlying substrate \citep{Valderrabano2003}, to simplify the visualization of three-dimensional scroll wave dynamics \citep{Fenton1998, Clayton2006}, and to measure fluctuations in the complexity of the dynamics \citep{Zaritski2004}. 
In short, phase singularities are an elegant way to characterize high-frequency arrhythmias that involve reentrant vortex waves, such as ventricular fibrillation (VF) or atrial fibrillation (AF) \citep{Nattel2017}.

Various methods have been proposed to compute phase maps and phase singularities (PS). 
These methods have been applied to both simulations of VF \citep{Fenton1998, Bray2001, Clayton2006} and AF \citep{Hwang2016, Rodrigo2017}, as well as experimental data, including electrode recordings of VF \citep{Nash2006, Umapathy2010} and AF \citep{Kuklik2015, Podziemski2018, Abad2021} in humans, %
optical maps of the transmembrane potential during VF \citep{Gray1998, Christoph2018, IyerGray2001, BrayWikswo2002, Rogers2004} and AF \citep{Yamazaki2012, Guillem2016} in isolated hearts, 
optical maps of action potential spiral waves in cardiac cell cultures \citep{Bursac2004, Entcheva2006, Munoz2007, Umapathy2010, You2017},
and time-varying 3D maps of mechanical strain waves measured during VF in isolated hearts using ultrasound \citep{Christoph2018}.
However, phase maps and PS are prone to measurement artifacts and deficits caused by inadequate processing of the measurement data, particularly when the data is noisy or sparse \citep{You2017, Roney2017, Kuklik2017, King2017, Rodrigo2017, Roney2019}. 
Noise and motion artifacts are a frequent issue when analyzing optical mapping recordings \citep{Zou2002, Christoph2018b}. 
Electrode mapping, used in both basic research and the clinical setting, is limited by low spatial resolution, or sparsity, even with the use of multi-electrode arrays and 64-lead basket catheters.

Mapping fibrillatory wave phenomena at low resolutions can lead to misrepresentation of the underlying dynamics. 
For example, low resolution phase mapping has been shown to create false positive detections of PS \citep{You2017, Roney2017, Kuklik2017, King2017, Roney2019}, contributing to much uncertainty in the imaging-based diagnosis of AF, a field in which rotors remain a highly controversial concept \citep{Aronis2017,Nattel2017,Schotten2020}. %
Mapping of AF would greatly benefit from computational methods, which could account for low spatial resolution and produce reliable visualizations of electrical phenomena from sparse and noisy spatio-temporal electrical signals.

\begin{figure}[htb]
  \centering
  \includegraphics[clip, trim=0.0cm 0.0cm 0.0cm 0.0cm, width=0.45\textwidth]{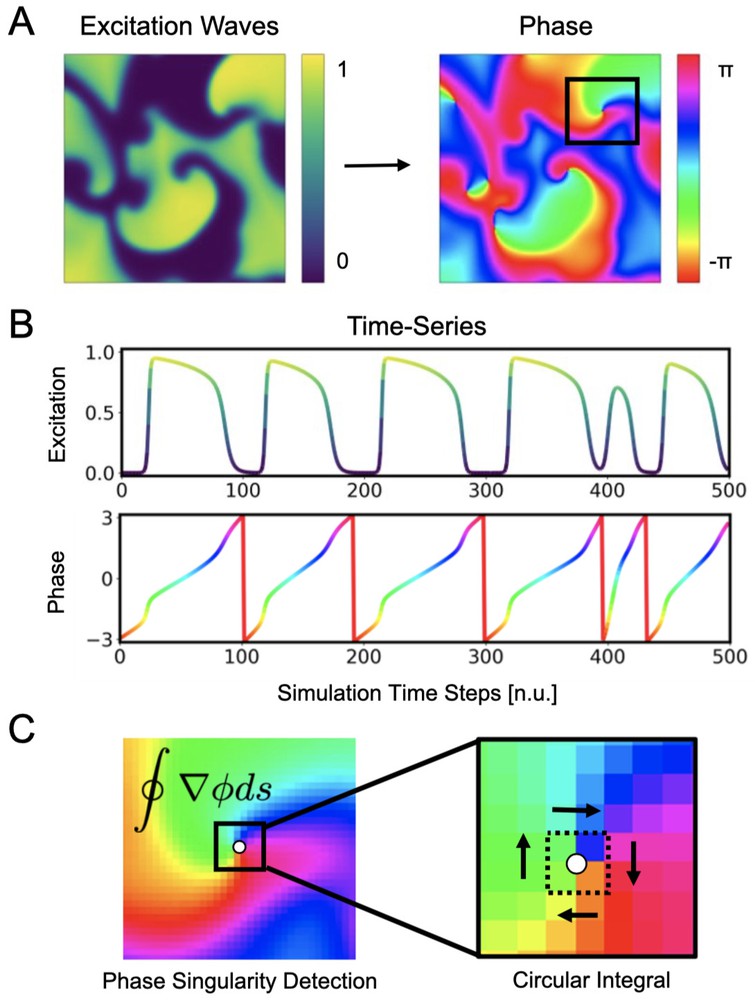}
  \caption{
  Cardiac electrical excitation wave pattern and conversion into corresponding phase map for the localization of rotor core positions or phase singularities (PS).
  A) Simulated electric spiral wave chaos pattern represented by transmembrane potential $V \in [0,1]$ (n.u., normalized units) and corresponding phase pattern with phase angle $\phi \in [-\pi,\pi]$.
  B) Time-series data showing a series of action potentials $V(t)$ and their representation as a phase signal $\phi(t)$ computed using the Hilbert transform.
  C) Classical phase singularity (PS) detection \citep{IyerGray2001} using a circular integral ($2 \stimes 2$ kernel) for the localization of spiral cores. Here the method is used to generate PS training data for deep learning-based PS detection, see Fig.~\ref{fig:Figure02}. Detailed field of view of region highlighted by black box in A).
  }
  \label{fig:Figure01Hilbert}
\end{figure}

\begin{figure*}[htb]
  \centering
  \includegraphics[clip, trim=0.0cm 0.0cm 0.0cm 0.0cm, width=0.72\textwidth]{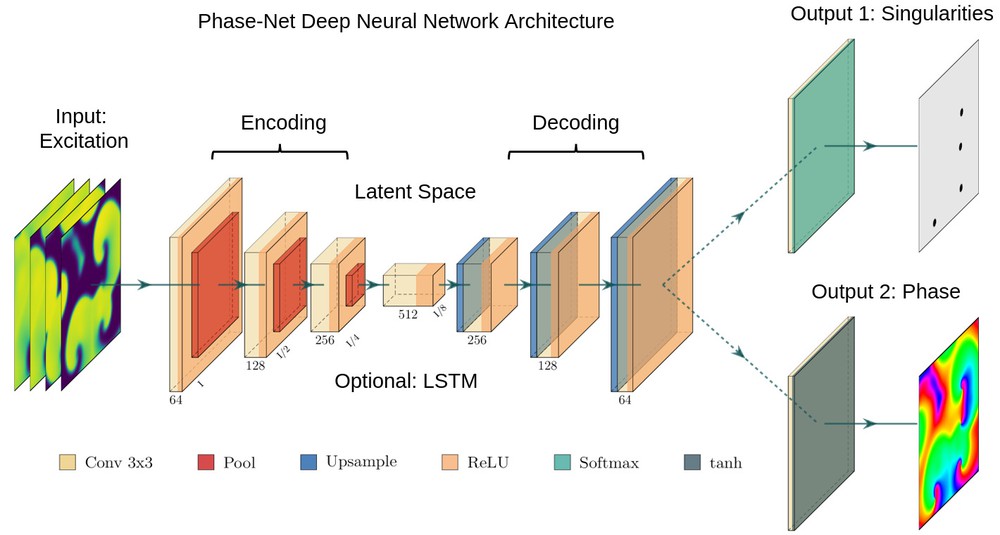}
  \caption{
    Deep convolutional neural network (CNN) with encoding stage, latent space and decoding stage for the computation of either phase maps or phase singularities (PS) from spatio-temporal maps of electrical excitation. Excitation, phase and PS data is used to train the two neural networks, which accordingly learn to translate a short sequence of excitation maps into a corresponding phase map or PS locations.
    After training, the networks can predict phase maps and PS positions from arbitrary unseen excitation data. 
    We used either i) a plain convolutional encoder-decoder network architecture, ii) a U-Net variant, or iii) a variant with a long short-term memory (LSTM) neural network module integrated into the latent space.
    The phase values are trigonometrically encoded as $x$- and $y$-components, see also Fig.~\ref{fig:FigureEncoding}.
  }
  \label{fig:Figure02}
\end{figure*}

In this study, we demonstrate that deep convolutional neural networks (CNNs) can be used to compute phase maps and phase singularities from short spatio-temporal sequences of electrical excitation wave patterns, even if these patterns are very sparse and very noisy.
We use variations of two-stage encoder-decoder CNNs with an encoding stage, a latent space, and a decoding stage, see Fig.~\ref{fig:Figure02}.
The neural network associates electrical excitation wave patterns with phase maps and phase singularity (PS) positions during a training procedure. After training, it is subsequently able to translate electrical excitation wave patterns into phase maps and PS when applied to new, previously unseen data. 
We tested a modified version of the neural network with an integrated convolutional long short-term memory (LSTM) module in the latent space of the original encoder-decoder architecture. 
Regardless of the particular architecture, the network was able to predict phase maps and PS in both experimental and synthetic data robustly and with high accuracy. 
When presented with sparse electrical data from a short temporal sequence of only 1-5 snapshots of electrical activity, the network maintained a robust accuracy level, even in the presence of strong noise. 
The approach may supersede more classical approaches due to its efficiency, its robustness against noise, and its ability to inter- and extrapolate missing measurement data with only minimal spatial and temporal information.

\quad
\subsection{Phase Mapping and Phase Singularity Detection Techniques}

\begin{figure}[htb]
  \centering
  \includegraphics[clip, trim=0.0cm 0.0cm 0.0cm 0.0cm, width=0.4\textwidth]{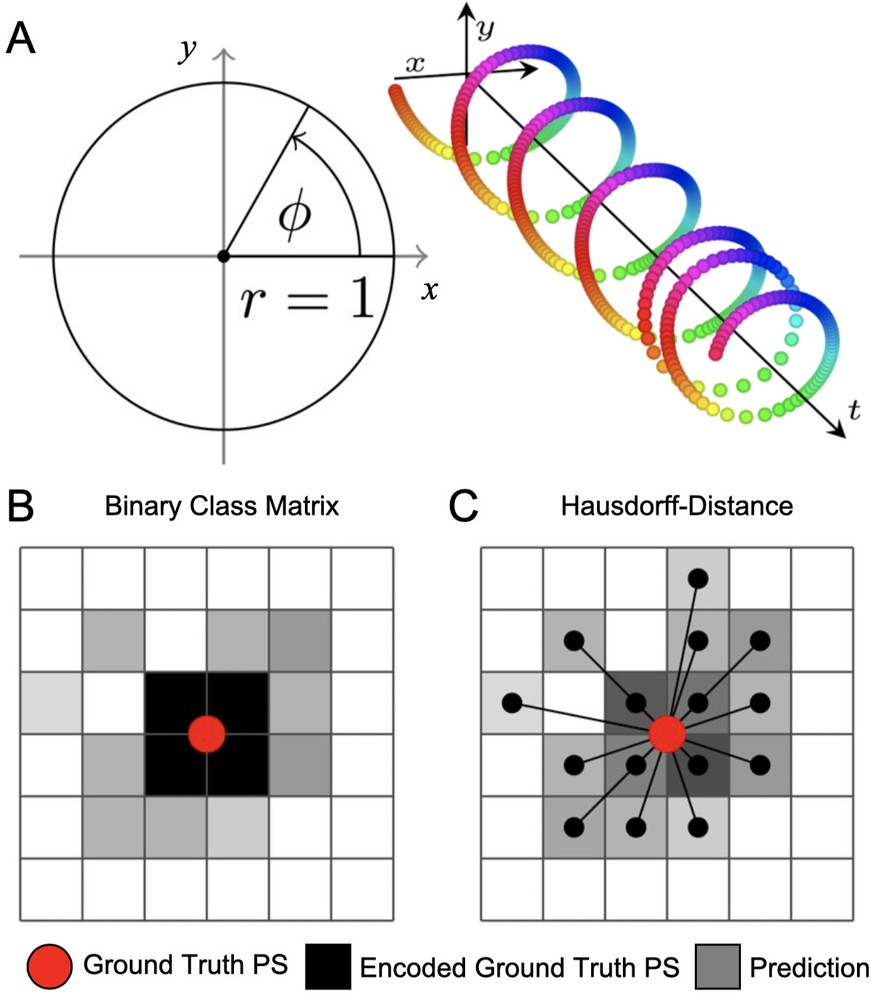}
  \caption{
 {Encoding of phase and phase singularities (PS) for deep learning.
 A) Instead of estimating the phase angle $\hat{\phi}$ directly, the phase prediction network produces two numbers $\hat{c}, \hat{s} \in [-1, 1]$ for each pixel as output.
 During training, these are compared against the trigonometric encoding of the phase angle $\phi\rightarrow(\cos(\phi), \sin(\phi))$. We define the predicted phase as $\hat{\phi} := \text{arctan2}(\hat{s}, \hat{c})$.
 B) Binary class matrix-encoding (or 'one-hot' encoding) of PS positions. The loss function (categorical cross-entropy) measures the difference between target PS (red) encoded as a $2{\times}2$ kernel (black) and predicted PS given as pixels (gray) with subsequent thresholding. Information about the distance between ground truth and predicted PS is not available to the minimization process during training.
 C) Coordinate-based encoding of PS positions. The loss function is based on the weighted Hausdorff-distance between ground truth PS coordinates and predicted PS positions given as pixel positions.
 }
 } 
  \label{fig:FigureEncoding}
\end{figure}

Phase maps and phase singularities (PS) have been used to characterize cardiac fibrillation for over 30 years \citep{Winfree1989}, and various methods were introduced to compute PS either directly or indirectly, see Fig.~\ref{fig:Figure01Hilbert}C). %
In computer simulations, the computation of a phase state or PS is straight-forward as the dynamic variables from the equations describing the local electrical state are readily available in the simulation and can be used to define a phase angle instantaneously \citep{Krinsky1992}.
For instance, with $V$ and $r$ for electrical excitation and refraction, respectively, see eqs. (\ref{eq:modelV})-(\ref{eq:modelr}), the phase angle can be defined as $\phi = \text{arctan2}( V , r)$, {see also Fig.~\ref{fig:FigureEncoding}A)}. 
Likewise, level-set methods using isocontour lines of two dynamic variables, such as $V$ and $r$, can be used to locate PS directly as the intersection points of these isocontours \citep{Barkley1990}. 
However, with experimental data, there is typically only one measured variable, such as the transmembrane voltage or an electrogram, and it is accordingly not possible to define a phase without additional temporal information. 
With experimental data, it becomes necessary to construct a phase signal $\phi(t)$ from a single measured time-series $V(t)$ using techniques such as i) delay embedding \citep{Gray1998}:
\begin{eqnarray}
  \phi(t) & = & \text{arctan2}(V(t),V(t+\tau)))
\end{eqnarray}
with an embedding delay $\tau$, typically defined as $\sim 1/4$ of the average cycle length or the first zero-crossing of the auto-correlation function, or ii) the Hilbert transform $\mathcal{H}(t)$, which generates the complex analytical signal of a periodic signal from which in turn the phase 
\begin{eqnarray}
  \phi(t) & = & \text{Re}( \mathcal{H}(t) )
\end{eqnarray}
can be derived \citep{BrayWikswo2002}.
The most intuitive approach to compute a time-dependent phase signal $\phi(t)$ of a sequence of action potentials is to detect the upstrokes of two subsequent action potentials and to define a piecewise linear continuous function $\phi_L (t)$, which linearly interpolates the phase angle from $-\pi$ to $\pi$ between the two upstrokes.
The Hilbert transform generates a phase signal $\phi_{\mathcal{H}}(t)$ which is very similar to the linearly interpolated phase signal $\phi_L(t)$, see Fig.~\ref{fig:Figure01Hilbert}B).

Phase singularities can then be calculated, see Fig.~\ref{fig:Figure01Hilbert}C), by using the circular line integral method developed by \citet{IyerGray2001} summing the gradient of the phase along a closed circular path $s$ around a point $\vec{x} = (x,y)$ in the phase plane:
\begin{eqnarray}
  \label{eq:lineintegral}
  \oint \nabla \phi (x,y;t) \text{d}s & = & \pm 2 \pi
\end{eqnarray}
If the circular path is sufficiently small (typically around $2 \stimes 2$ pixels), the integral yields $\pm 2 \pi$ when the line integral encloses a phase singularity (the sign indicates chirality), or $0$ if it does not enclose a phase singularity.
As the line integral method calculates the spatial gradient of the phase, it is very sensitive to noise and requires continuous and smooth phase maps.
Therefore, much prior work has focused on improving the robustness of phase mapping and PS detection methods under more realistic conditions, e.g with noise or other artifacts that typically occur with, for instance, contact electrode measurements.
\citet{Zou2002} further refined the line integral method using convolutions and image analysis.
\citet{Kuklik2015} introduced sinusoidal recomposition to remove undesired high-frequency components during the computation of phase signals using the Hilbert transform.
In contrast to the line integration method, \citet{Tomii2016} proposed computing the phase variance to locate PS.
Similarly, \citet{Lee2016} introduced a so-called "location-centric" method to locate PS, the method only requiring temporal information about the voltage at the core.
\citet{Li2018} introduced a Jacobian-determinant method using delay embedding for identifying PS also without explicitly computing a phase.
\citet{Marcotte2017} and Gurevich et al. introduced level-set methods to compute PS in noisy conditions and demonstrated the robustness of the approach with VF optical mapping data \citep{Gurevich2017,Gurevich2019}.
\citet{Vandersickel2019} proposed to use graph theory to detect rotors and focal patterns from arbitrarily positioned measurement sites.
\citet{Mulimani2020} used CNNs to detect the core regions of simulated spiral waves using a CNN-based classification approach and discriminating sub-regions containing spiral wave tips from areas exhibiting other dynamics, and consequently generated low-resolution heat maps indicating the likely and approximate core regions of spiral waves.
Very similarly, \citet{Alhusseini2020} used CNNs to classify and discriminate rotational and non-rotational tiles in maps of AF acquired with basket catheter electrode mapping.
Lastly, \citet{Li2020} provided a comparison of 4 different PS detection algorithms applied to AF and found that results can vary significantly.

\section{Methods}
We developed a two-stage deep convolutional neural network (CNN) with encoder and decoder architecture and trained the network with pairs of two-dimensional maps showing electrical excitation wave patterns and corresponding 'ground truth' phase maps and phase singularity (PS) locations. 
The training was performed with both simulation data and experimental data, which was obtained in voltage-sensitive optical mapping experiments {in two different species} during ventricular fibrillation (VF).
After training, the network was applied to new data and used to predict phase maps or the positions of PS from 'unseen' excitation wave patterns.

\quad
\subsection{Neural Network Architecture}\label{sec:methods:networkarchitecture}

{The architecture of our neural networks comprises an encoding stage, a latent space, and a decoding stage, see Fig.~\ref{fig:Figure02}.}
The neural networks are designed to translate an arbitrary two-dimensional electrical excitation wave pattern or a short sequence of two-dimensional excitation wave patterns into {either} a corresponding two-dimensional phase map, or predict the positions of phase singular points (PS) in the electrical maps.
{We developed {three} phase map prediction neural network models M1, M2 and M3, and two different PS prediction neural network models M1A and M1B which are based on M1.
The three phase map prediction models are a basic encoder-decoder CNN version M1, an LSTM-version M2 and a U-Net version M3, see below for details.
The difference between models M1A and M1B is mainly the associated loss function and the encoding of the ground truth PS. 
M1A uses a pixel-wise cross-entropy loss which does not account for the distance between predicted PS locations and ground truth PS unless they overlap, whereas M1B uses a loss function based on the distance between predicted and ground truth PS locations.}

{The phase map prediction neural networks are trained with excitation wave patterns as input and a two-dimensional trigonometric encoding of the phase map as target, see Fig.~\ref{fig:FigureEncoding}A).
The trigonometric encoding eliminates the discontinuity of a linear encoding of the cyclic phase $\phi$ by encoding the value onto a two-dimensional unit circle: $\phi \rightarrow (\cos(\phi), \sin(\phi)) =: (c, s)$.
Therefore, the two phase mapping CNNs have a two-dimensional layer with two channels as output, which are estimates of the sine $\hat{s}$ and cosine $\hat{c}$ of the phase angle $\phi$.
The predicted phase $\hat{\phi}$ is decoded as $\hat{\phi} := \text{arctan2}(\hat{s}, \hat{c})$.
We use the hyperbolic tangent function as activation function in the last layer of the phase mapping CNNs {to ensure that $\hat{c}, \hat{s}\in[-1, 1]$.
All models are based on a convolutional encoder-decoder architecture, see Fig.~\ref{fig:Figure02}. 
However, whereas model M1 uses a two-dimensional convolutional layer in the latent space, the latent space of model M2 is a two-dimensional convolutional long short-term memory (LSTM) neural network layer \citep{Hochreiter1997,Shi2015}, and model M3 is based on the generic convolutional architecture of model M1, but includes long skip connections at each maxpooling/upsampling step, similar to U-Net \citep{Ronneberger2015}.
In all models the encoder- and decoder-stage consist of three two-dimensional convolutional layers, each followed by a batch normalization layer \citep{Ioffe2015}, rectified linear unit (ReLU) activation layer \citep{Nair2010}, and a maxpooling or upsampling layer.
The convolutional layers use 64, 128, and 256 kernels in the encoding stage, 512 kernels in the latent space, and 256, 128, and 64 kernels in the decoder stage.
The phase prediction models use the mean squared error as loss function.}}

{The two PS prediction neural networks M1A and M1B are trained with excitation wave patterns as input and either i) a dense binary class matrix representation of PS positions or ii) coordinates of PS positions as target, respectively, see Fig.~\ref{fig:FigureEncoding}B,C).
The ground truth PS are located -- by construction, see Fig.~\ref{fig:Figure01Hilbert}C) -- in the center of a $2{\times}2$ kernel. 
With model M1A we set $1$ as target for all four neighboring pixels of a PS and $0$ for all other pixels.
While it is possible to train directly on such an encoding with a binary cross-entropy loss function, we achieved better accuracies when using a categorical encoding of the target image as a $128{\times}128{\times}2$ class matrix, where in the first channel all non-PS pixels are valued $1$ and in the second channel all $2{\times}2$ PS pixels are $1$ and $0$ otherwise. 
Accordingly, model M1A uses two output layers with a softmax activation function, and categorical cross-entropy as a loss function during training. 
Note that the loss corresponds to a pixel-wise loss, which does not take into account distances between ground-truth and approximated  PS positions.
With model M1B the target PS are encoded directly as a list of two-dimensional $(x,y)$-coordinates of PS positions and the loss function uses a weighted Hausdorff-distance with the paramter $\alpha=-3$ between the target PS and predicted pixel distributions approximating PS positions, which was introduced by \citet{Ribera2019} for the deep learning-based localization of objects, see illustration in Fig.~\ref{fig:FigureEncoding}C).
Note that the loss function includes information about spatial distances between ground-truth and approximated PS during training.
Model M1B comprises one output layer with a sigmoid activation function and we used a threshold of $0.5$ to obtain a binary PS prediction image.
For both models M1A and M1B the predicted PS positions are computed as sub-pixel precise PS locations from the center of each connected object in the binary PS prediction image.
Two pixels are connected (belong to the same object), if both are $1$ and when their edges or corners are adjacent.
}

\begin{figure}[htb]
  \centering
  \includegraphics[clip, trim=0.0cm 0.0cm 0.0cm 0.0cm, width=0.8\textwidth]{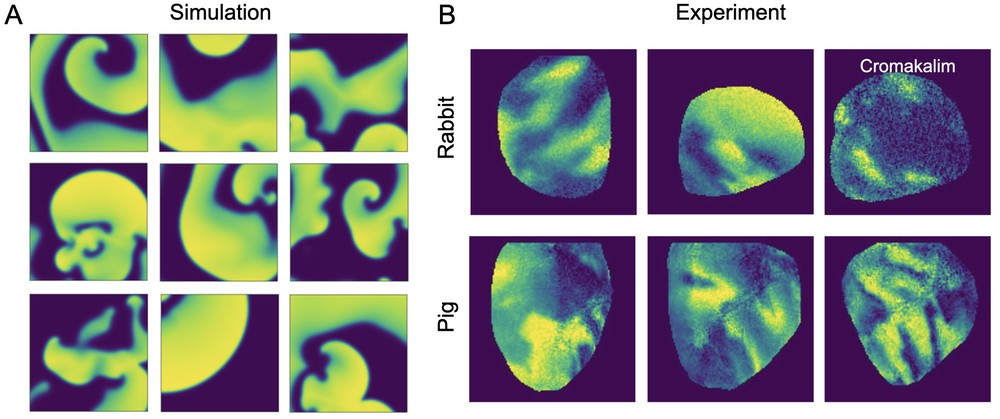}
  \caption{
  {Simulated and experimental training data. Each training dataset includes $20{,}000$ samples. 
  A) Random snapshots of simulated electrical spiral wave chaos. The dynamics are diverse and include both chaotic and laminar episodes with both spiral and plane waves and include longer and shorter wavelengths and faster and slower conduction speeds, respectively. The simulated training data was further noisified and/or sparsified, see Fig.~\ref{fig:FigurePreprocessing}B) and \ref{fig:FigureResults2Phase}A).
  B) Experimental training data generated in voltage-sensitive optical mapping experiments during ventricular fibrillation (VF) in rabbit (top) and pig (bottom) hearts. The rabbit data contains about $50\%$ VF episodes with Cromakalim and $50\%$ without. Therefore, both datasets include shorter and longer action potential wavelengths, as well as faster and slower and more and less complex dynamics, respectively.}
  }
  \label{fig:FigureTrainingData}
\end{figure}

All network models analyze either a single, static two-dimensional excitation wave pattern or a short sequence of up to $10$ excitation wave patterns as input. The patterns consist of consecutive snapshots of the activity sampled at the current time step $t$ and at equidistant time intervals at previous time steps, see also section \ref{sec:methods:trainingdata}.
Note that, if we refer to 'video images / frames / excitation patterns' or 'samples', each of these samples may refer to a single or a short series of $2-10$ two-dimensional excitation patterns.
For model M1 {and M3} the excitation wave patterns are represented as input channels, while for model M2 each {temporal} excitation wave pattern is processed separately in the neural network {as the LSTM is a recurrent neural network}.
All neural network models were implemented in Tensorflow \citep{tensorflow2015-whitepaper} version 2.6.0.

\quad
\subsection{Training Data Generation}\label{sec:methods:trainingdata}
We generated synthetic training data using a phenomenological computer model of cardiac electrophysiology \citep{AlievPanfilov1996}.
In short, nonlinear waves of electrical excitation and refractoriness were modeled using partial differential equations and an Euler finite differences numerical integration scheme:
\begin{eqnarray} 
\label{eq:modelV}
\frac{\partial V}{\partial t} & = & \nabla^2 V - k V (V-a) (V-1) - V r \\
\label{eq:modelr}
\frac{\partial r}{\partial t} & = & \epsilon(V,r)(k V(a+1-V)-r)
\end{eqnarray}
Here, $V$ and $r$ are dimensionless, normalized dynamic variables for electrical excitation (voltage) and refractoriness, respectively.
Together with the isotropic diffusive term $\nabla^2 V = \nabla \cdot (D \nabla V)$ with the diffusion constant $D=1.0$ in eq. (\ref{eq:modelV}), the model produces nonlinear waves of electrical excitation and the term $\epsilon(V,r) = \epsilon_0 + \mu_1 r / (V+\mu_2)$ in eq. (\ref{eq:modelr}) and electrical parameters $k$, $a$, $\epsilon_0$, $\mu_1$ and $\mu_2$ influence properties of the excitation waves.
The size of the two-dimensional simulation domain was $200 \stimes 200$ cells/pixels.
The parameters were set to $a=0.09$, $k=8.2$, $\epsilon_0=0.01$, $\mu_1=0.07$, $\mu_2=0.3$ and spiral wave chaos was initiated by applying a series of point stimulations in random locations.
With the chosen parameters the dynamics exhibit both chaotic spiral wave and more laminar wave dynamics with strong fluctuations in the complexity of the wave patterns, see Figs.~\ref{fig:FigureTrainingData}A) and \ref{fig:FigureResults1}F) and Supplementary Video 1.
We generated $20$ episodes with a series of $2{,}500$ snapshots of the dynamics in each episode. 
Fig.~\ref{fig:Figure01Hilbert}A) shows an example of such a snapshot.
The $2{,}500$ snapshots show about $25$ spiral wave rotations. 
Correspondingly, one spiral rotation is resolved by about $100$ snapshots.
Note that in the simulation the dynamics are resolved at a $10\times$ higher temporal resolution than in the series of snapshots, because we stored a snapshot only in every 10th simulation time step.
In total, we obtained $50{,}000$ snapshots, from which we then created $20{,}000$ training samples, see Fig.~\ref{fig:FigureTrainingData}A), where one training sample comprises a short sequence of snapshots with up to $10$ images of the excitation. 
Within the sequence, the first snapshot, denoted with $t_0$, corresponds to the snapshot at time $t$ in the video. 
The training is performed with the corresponding ground truth phase maps and PS obtained at this time step $t$ and, correspondingly, the network also predicts a phase map or PS at time $t$.
The other snapshots in each sample correspond to snapshots showing the dynamics at previous time steps $t_{-1}$, $t_{-2}$ etc., where $t_{-i}=t_0- i \cdot \tau$ with $i=1,\ldots,N_t$ and $N_t$ is the number of snapshots in the sample and $\tau$ is the temporal sampling distance between the frames over parts of the previous period.
The parameters $N_t$ and $\tau$ are discussed in more detail in section \ref{sec:results:spatiotemporal} and in Fig.~\ref{fig:FigureSampling}.
The training samples were shuffled in time, while the temporal sequence within each sample was kept in its original order.
We generated test data for evaluation that was not used during training by simulating $5{,}000$ snapshots separately using the same electrical parameters and generating samples with the same parameters $N_t$ and $\tau$ for testing purposes.
We computed ground truth phase maps from the original series of excitation snapshots before shuffling using the Hilbert transform \citep{BrayWikswo2002} and computed ground truth PS using the \citet{IyerGray2001} line integral method, as shown in Fig.~\ref{fig:Figure01Hilbert}B,C). 
To simulate noisy excitation wave data, we added noise to the training data, see Fig. \ref{fig:FigureNoiseVsSparsity}.
The Gaussian white noise was added to the individual pixels independently in each frame and independently over time ($\sigma = 0.1, 0.2, \ldots , 0.8$ states the standard deviation of the noise).
Each excitation snapshot was optionally additionally sparsified by setting masked excitation values to $0$.

\begin{figure}[htb]
  \centering
  \includegraphics[clip, trim=0.0cm 0.0cm 0.0cm 0.0cm, width=0.45\textwidth]{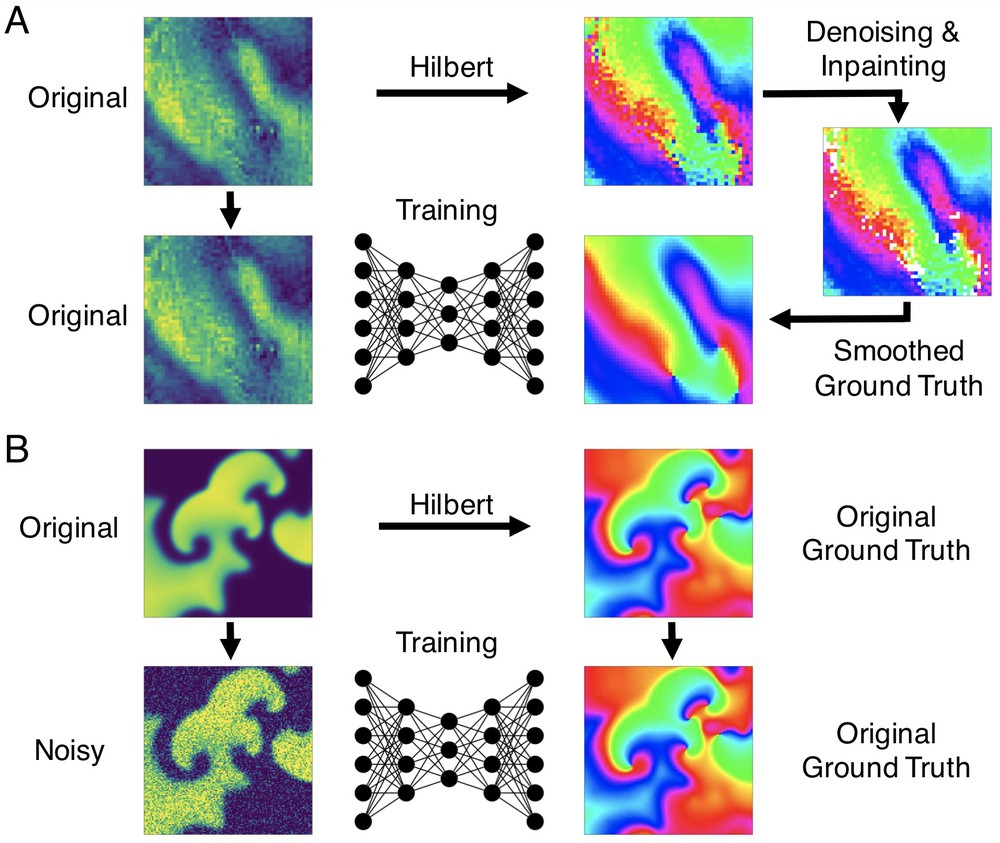}
  \caption{
  Preprocessing of data for training of deep learning algorithm for computation of phase maps and phase singularities.
  A) Preprocessing of noisy optical mapping data: computation of noisy phase maps from raw noisy optical maps using the Hilbert transform. Subsequent denoising using outlier removal, inpainting and smoothing to obtain ground truth phase maps for training.
  B) Preprocessing of simulation data: computation of ground truth phase maps directly from smooth simulated excitation wave patterns and training with noisy (or sparsified) excitation data.
  }
  \label{fig:FigurePreprocessing}
\end{figure}

We generated experimental training data using high-speed video data obtained in optical mapping experiments with voltage-sensitive fluorescent dyes (Di-4-ANEPPS).
Imaging was performed during VF in isolated {rabbit ($N=2$) and porcine ($N=5$) hearts} at acquisition speeds of {$500\,\text{fps}$, respectively, using a Teledyne Photometrics Evolve camera ($128{\times}128$ pixels).  
The rabbit data included $6$ recordings with $4$ different views and more than $25{,}000$ video frames in total.
The pig data included $10$ recordings with $8$ different views and more than $100{,}000$ video frames in total. 
About half of the rabbit data shows VF episodes with the potassium channel opener Cromakalim, which typically reduces the action potential duration and accelerates VF dynamics.} 
The raw optical mapping videos were pixel-wise normalized in time using a sliding-window normalization (window size $100$ - $120$ frames). 
We used the Hilbert transform to {compute phase maps of the pixel-wise normalized optical maps, the phase maps were subsequently denoised and smoothed, see Fig.~\ref{fig:FigurePreprocessing}A) and section \ref{sec:methods:interpolation}, to obtain ground truth phase maps.}
These ground truth phase maps were then used to compute ground truth PS using the circular integral method as with the simulation data.
{$20{,}000$ samples of the pixel-wise normalized noisy versions of the voltage-sensitive optical maps (without spatio-temporal smoothing), ground truth phase maps, and PS were used as training dataset for each species, see Fig.~\ref{fig:FigurePreprocessing}A).}
{The test datasets consisted of $5{,}000$ samples, which were derived from 1-2 separate recordings, which were left out of the training dataset. Each training or test sample corresponds to a short series (10 frames) of voltage-sensitive maps showing action potential wave dynamics in analogy to the simulation data.}
{The experimental samples were masked with masks outlining the shape of the heart. Pixels outside of the mask were set to $0$. 
The same masks were also applied to simulated data, see Supplementary Video 2.}

\quad
\subsection{Training Procedure}\label{sec:methods:training}

Using the experimental and simulated data described in section \ref{sec:methods:trainingdata}, we generated training datasets consisting of corresponding two-dimensional electrical excitation wave data and phase maps as well as $(x,y)$ positions of PS in these maps.
The simulated data was resized from $200{\times}200$ pixels to $128{\times}128$ pixels to match the size of the experimental data.
{All predictions were performed on a separate dataset, which was not part of the training. 
The predictions in Figs.~\ref{fig:FigureResults1}-\ref{fig:FigureSampling} were only performed on 'unseen' data, which the neural network was not exposed to during training.} 
A fraction of $5\%$ of the samples of the training datasets were used for validation during training.
The networks were trained with a batch size of $32$ using the Adam \citep{Adam} optimizer with a learning rate of $0.001$.
All models were typically trained for $10$ to $15$ epochs on data including $20{,}000$ frames or samples, if not stated otherwise.

\quad
\subsection{Phase Mapping and Rotor Localization Accuracy}\label{sec:methods:accuracy}
{
The phase prediction accuracy was determined by calculating the angular accuracy, $1- \left\langle |\Delta \phi| \right\rangle / \pi$, where $|\Delta \phi|$ is the minimum absolute angle difference between the predicted phase $\hat{\phi}_i(x,y)$ and the ground truth phase $\phi_i(x,y)$. The average absolute angle difference
\begin{eqnarray}
  \left\langle |\Delta \phi| \right\rangle & = & \frac{1}{N\cdot N_{\text{pixels}}} \sum_{i, x, y} |\Delta \phi_i(x, y)|
  \label{eq:error}
\end{eqnarray}
is evaluated over all $N_{\text{pixels}}$ pixels $(x, y)$ in all $N$ test samples $i$.
All uncertainties of the phase prediction accuracies stated throughout this study correspond to the standard deviation of the angular accuracy over all $N_{\text{pixels}}$ pixels in all $N$ samples in the entire testing dataset.}
The PS prediction accuracy was evaluated with the precision, recall, and F-score based on the number of true positive $tp$, false positive $fp$ and false negative $fn$ PS predictions{, as well as the mean absolute error of the number of predicted PS and the mean average Hausdorff distance}.
A true positive estimated PS position is counted if any estimated PS location is within at most $r$ pixels from the ground truth PS.
A false positive is counted if no ground truth PS is located within a distance of $r$ from the estimated PS position.
A false negative is counted if a ground truth PS does not have any estimated PS within a distance of at most $r$.
We chose $r=3$ pixels, see also Fig.~\ref{fig:FigureResults1}E).
{We note that this definition is biased in favor of the prediction when two PS are predicted within $r$ pixels of a single ground truth PS, as both predicted PS will be counted as true positive.
However, by construction of the prediction method (see Fig.~\ref{fig:FigureEncoding} and section \ref{sec:methods:networkarchitecture}) this case occurs only very rarely.
E.g., for none of the models presented in Table~\ref{tab:pseval} did this situation occur for more than 15 PS out of a total of $\sim 17{,}000$ predicted PS. The bias in favor of the model is thus negligible for the precision, recall and F-score.}
Precision is $tp / (tp + fp)$, the proportion of estimated PS locations that are close enough to a ground truth PS location.
Recall is $tp / (tp + fn)$, the proportion of the true phase singularities the neural network is able to detect.
The F-score is the harmonic mean of precision and recall:
\begin{eqnarray}
  \text{F-score} = 2 \cdot \frac{\text{Precision}\cdot\text{Recall}}{\text{Precision}+\text{Recall}}
\end{eqnarray}
{
Additionally, we compute the mean absolute error (MAE) of the number of predicted PS
\begin{eqnarray}
  \text{MAE}&=&\frac{1}{N} \sum_{i=1}^N \left|\hat{n}_i-n_i\right|
\end{eqnarray}
where $N$ is the number of dataset samples, $n_i$ is the number ground truth PS in the $i$-th sample, and $\hat{n}_i$ is the number of predicted PS for the sample.
The average Hausdorff distance $d_\text{AHD}$ measures the distance between two point sets $X$ and $Y$:
\begin{eqnarray}
  d_\text{AHD}(X, Y) =\frac{1}{2}\left(\frac{1}{|X|} \sum_{\vec{x} \in X} \min_{\vec{y}\in Y} \|\vec{x}-\vec{y}\| + \frac{1}{|Y|} \sum_{\vec{y}\in Y} \min_{\vec{x} \in X} \|\vec{x}-\vec{y}\|\right)
\end{eqnarray}
where $|X|$ and $|Y|$ are the number of points in $X$ and $Y$ respectively and $\|{{}\cdot{}}\|$ is the Euclidean distance.
We report the mean average Hausdorff distance for PS predictions
\begin{eqnarray}
  \text{MAHD} = \frac{1}{N} \sum_{i=1}^N d_{\text{AHD}}(S_i, \hat{S}_i)
\end{eqnarray}
where $S_i$ is the set of ground truth PS and $\hat{S}_i$ is the set of predicted PS for sample $i$. If either $S_i$ or $\hat{S}_i$ is empty and the other set is not empty we set $d_{\text{AHD}}(S_i, \hat{S}_i)$ to the image diagonal in pixels.
}

{
\subsection{Smoothing and Interpolation}
\label{sec:methods:interpolation}
To be able to compare the CNN-based phase predictions shown in Fig.~\ref{fig:FigureResults2Phase}B) with results obtained with a reference method, we reconstructed or enhanced the noisy and/or sparse phase maps shown in Fig.~\ref{fig:FigureResults2Phase}C) using kernel-based spatio-temporal outlier filter, inpainting and smoothing techniques.
The filtering techniques were also applied to experimental data, see Fig.~\ref{fig:FigurePreprocessing}A) and section \ref{sec:methods:trainingdata}.
The filtering is performed on trigonometrically encoded phase values, where each real-valued phase value in the video is converted into its complex decomposition:
\begin{eqnarray} 
\label{eq:encoding}
\phi (x,y;t) & \rightarrow & \cos{\phi(x,y;t)} + i \cdot \sin{\phi(x,y;t)}
\end{eqnarray}
Spatio-temporal kernels are then used to average the complex phase values in space and over time in small disk-shaped sub-regions $\mathcal{S}_{d,\Delta t}$ with diameter $d$ and with $\Delta t = 3$ at times $t-1$, $t$ and $t+1$. 
In order to remove outliers in the experimental data and the noisy simulated data, the Kuramoto order parameter $r(x, y; t)$ \citep{Kuramoto1984} was computed in every pixel at every time step:
\begin{eqnarray} 
\label{eq:orderparameter}
r \cdot e^{\phi}& = & \frac{1}{N} \sum \limits_i^N e^{i \phi_j}
\end{eqnarray}
where $j=1,...,N$ is the number of complex phase values within each kernel with diameter $d=5$ pixels and $\Delta t=3$. 
Phase values were considered outliers if $r<0.9$ and accordingly removed, as shown in Fig.~\ref{fig:FigurePreprocessing}A).
Missing phase values were replaced with phase values averaged from surrounding phase values within the spatio-temporal kernel, given that at least $30\%$ of the entries within the kernel were non-missing or valid phase entries. 
The process was repeated until the entire video was filled with valid phase entries.
Lastly, the denoised, inpainted phase maps were smoothed averaging all phase values within a small spatio-temporal kernel typically with $d=7$ and $\Delta t =3$, if not stated otherwise.
In Fig.~\ref{fig:FigureResults2Phase}, the noisy data was processed using the outlier and smoothing filters, the low resolution data was smoothed with $d=11$ pixels, the $8{\times}8$ large and small grid data was inpainted $7$ times with $d=11$ pixels, and the sparse grid data was inpainted $10$ times with $d=19$ pixels, all with $\Delta t=3$.
With the sparse data the denoising was performed after inpainting and before smoothing.}

\section{Results}

\begin{figure}[htb]
  \centering
  \includegraphics[clip, trim=0.0cm 0.0cm 0.0cm 0.0cm, width=0.46\textwidth]{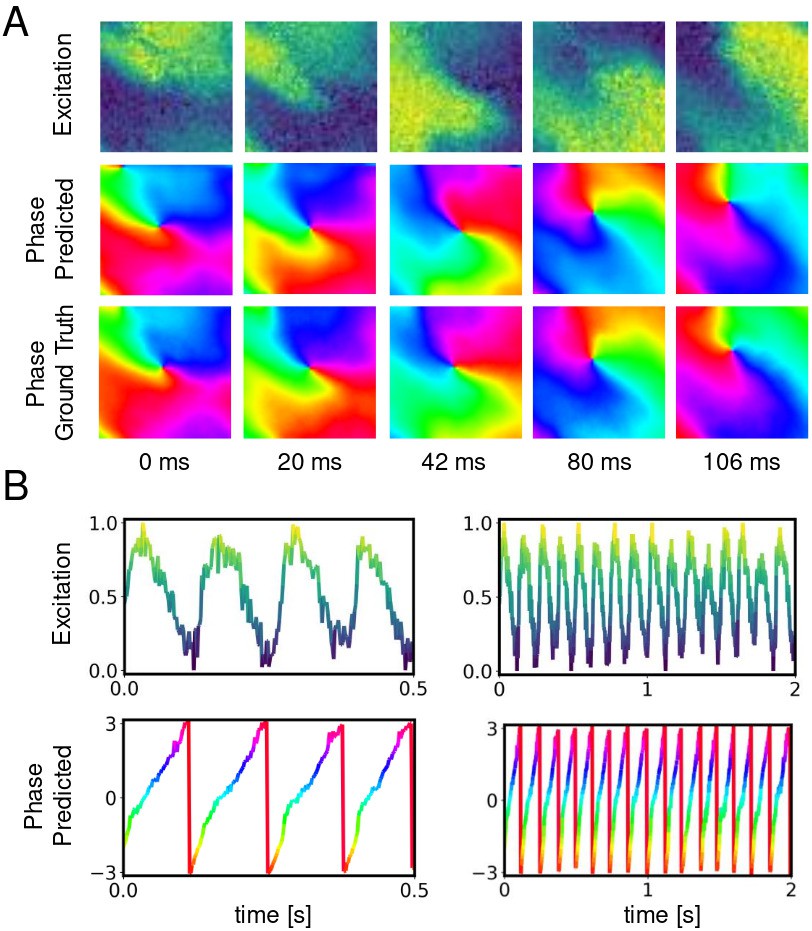}
  \caption{
  Deep neural network-based prediction of phase maps from optical maps measured using voltage-sensitive fluorescent dye Di-4-ANEPPS during ventricular fibrillation on surface of isolated heart.
 A) Optical maps of transmembrane voltage showing counter-clock-wise rotating action potential spiral vortex wave (normalized units [0,1], pixel-wise normalization, yellow: depolarized tissue, blue: refractory tissue). 
 Comparison of predicted (top) and ground truth (bottom) phase maps with high qualitative and quantitative agreement. The phase prediction accuracy is $97\% \pm 6\%$ and the predicted and ground truth phase maps are hard to distinguish. 
 The data was not seen by the network during training.
 B) Exemplary time-series from a single pixel showing transmembrane voltage $V(t)$ and predicted phase $\hat{\phi}(t)$, respectively.
  }
  \label{fig:FigureExperiments}
\end{figure}

\begin{figure}[htb]
  \centering
  \includegraphics[clip, trim=0.0cm 0.0cm 0.0cm 0.0cm, width=0.45\textwidth]{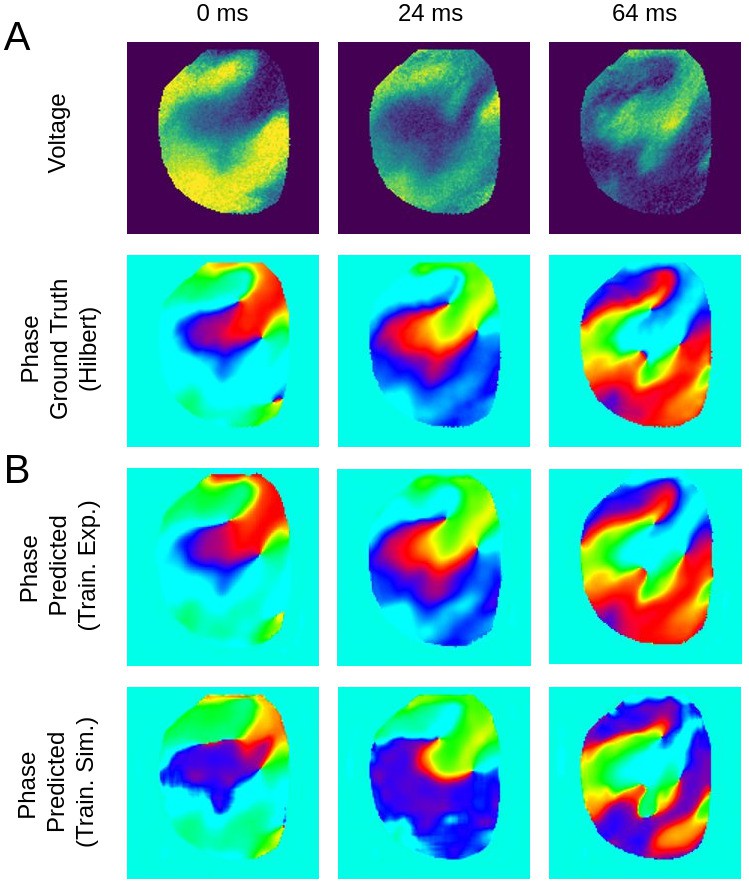}
  \caption{
  {Deep learning-based phase mapping of VF in rabbit heart with neural network trained on either experimental or simulation data.
  A) Voltage-sensitive normalized optical maps showing action potential vortex waves during VF on rabbit heart and corresponding ground-truth phase maps computed using the Hilbert transform. 
  B) Prediction of phase maps using neural network model M1 trained with either rabbit optical mapping data (top, data not seen during training) or solely simulated data (bottom) of excitation spiral wave chaos (noise $\sigma = 0$) as shown in Fig.~\ref{fig:FigureTrainingData}A) but masked as in Fig.~\ref{fig:FigureTrainingData}B).
  The phase prediction accuracy is $97 \pm 6 \%$ and $94 \pm 11 \%$ when training is performed with experimental or simulation data, respectively, see also Fig.~\ref{fig:FigureCrossTraining} for a comparison of prediction accuracies when training across different species.} 
  }
  \label{fig:FigureSimulationToExperiment}
\end{figure}

{We found that deep encoding-decoding convolutional neural networks (CNNs) can be used to compute phase maps and phase singularities (PS) from a short sequence of excitation wave patterns.
The prediction of phase maps can be performed robustly and accurately ($\sim 90-99\%$) with both experimental and simulated data, even with extremely noisy or sparse patterns, see Figs.~\ref{fig:FigureExperiments}-\ref{fig:FigureSampling} and Supplementary Videos 1 and 4-6.
Phase predictions remained accurate across different species, with models being trained on one species and then being successfully applied to another. 
Additionally, models that were trained solely on simulation data of VF could be applied to experimental data, see Figs.~\ref{fig:FigureSimulationToExperiment} and \ref{fig:FigureCrossTraining}.
PS can be predicted either directly from excitation wave patterns or indirectly by first predicting phase maps from excitation wave patterns and then computing PS in the predicted phase maps.
While in principle both direct and indirect PS prediction methods can determine the positions of PS very precisely {(F-scores of $\sim97\,\%$, see Table~\ref{tab:pseval})}, direct PS predictions are very sensitive to noise and sparsity. Indirect PS predictions are far more robust.
Accordingly, with the indirect PS prediction method we were able to locate PS in optical mapping recordings of VF sufficiently reliably and accurately, whereas with the direct PS prediction method this task was more challenging and produced only {moderately successful results, see Table~\ref{tab:PerformancePSPredictionRabbit}}.}


Figs.~\ref{fig:FigureExperiments} and \ref{fig:FigureSimulationToExperiment} and Supplementary Video 1 show predictions of phase maps when the neural network analyzes voltage-sensitive optical mapping videos showing action potential spiral vortex waves during ventricular fibrillation (VF) on the surface of {rabbit and porcine hearts}. 
Fig.~\ref{fig:FigureExperiments}A) shows raw pixel-wise normalized optical maps with a counter-clock-wise rotating action potential spiral vortex wave {on the ventricular surface of an isolated pig heart} (close-up, $48{\times}48$ pixels cutout from original video image).
The action potential rotor performs one rotation in about $110\,\text{ms}$.
The phase maps in the second and third row in panel A) show the predicted phase maps $\hat{\phi}$ obtained with model M1 and ground truth phase maps $\phi$, respectively. 
The action potential rotor is characterized by a pinwheel pattern in the phase maps, and the rotational core or PS is indicated by lines of equal phase which merge at the center of the pinwheel pattern.
Predicted and ground truth phase maps are visually almost indistinguishable and exhibit only minor differences.
The data was not seen by the neural network previously during training.
The predicted phase maps are smooth even though the optical maps showing the action potential wave patterns are noisy. 
The neural network is able to predict more complicated wave patterns with multiple rotors or phase singularities, see Fig.~\ref{fig:FigureSimulationToExperiment} and Supplementary Videos 1 {and 3}.
{The upper row in Fig.~\ref{fig:FigureSimulationToExperiment}B) shows phase map predictions of an action potential figure-of-eight reentry pattern on the ventricular surface of a rabbit heart during VF. 
The predicted and ground truth phase maps, shown in Fig.~\ref{fig:FigureSimulationToExperiment}A), can only be distinguished from each other upon close inspection.}
Analyzing a short sequence of $10$ optical maps, the neural network provides phase map predictions, which are very accurate and sufficiently smooth in both space and over time, and the predictions can be retrieved in real-time at an acquisition speed of $500\,\text{fps}$.
Fig.~\ref{fig:FigureExperiments}B) shows an optical trace of a series of action potentials and the corresponding time-series of the predicted phase, which was obtained from the sequence of predicted phase maps in A) using model M1. 
Even though each phase map was predicted independently at each time step, the time-course of the predicted phase signal $\hat{\phi} (t)$ is relatively smooth, see Supplementary Videos for an impression of the temporal smoothness of the predictions.  
{On average, the accuracy of the phase prediction with model architecture M1 is $97\% \pm 8\%$ or $98\% \pm 6\%$ in terms of angular accuracy, if the model was trained on pig data and is evaluated on pig data or, alternatively, trained on rabbit data and evaluated on rabbit data (evaluation on $\sim 5{,}000$ frames that were not part of the training data), respectively.
We did not find a significant difference in the accuracy between models M1, the LSTM model M2, or the U-Net style model M3. For instance, when trained and evaluated on pig data, the angular accuracy for the phase prediction was $96.5\%\pm7.9\%$ for M1, $96.1\%\pm8.1\%$ for M2 and $96.7\%\pm7.8\%$ for model M3.}

\quad
\subsection{Phase Prediction across Species and Dynamical Regimes}\label{sec:crosstraining}

\begin{figure}[htb]
  \centering
  \includegraphics[width=0.45\textwidth]{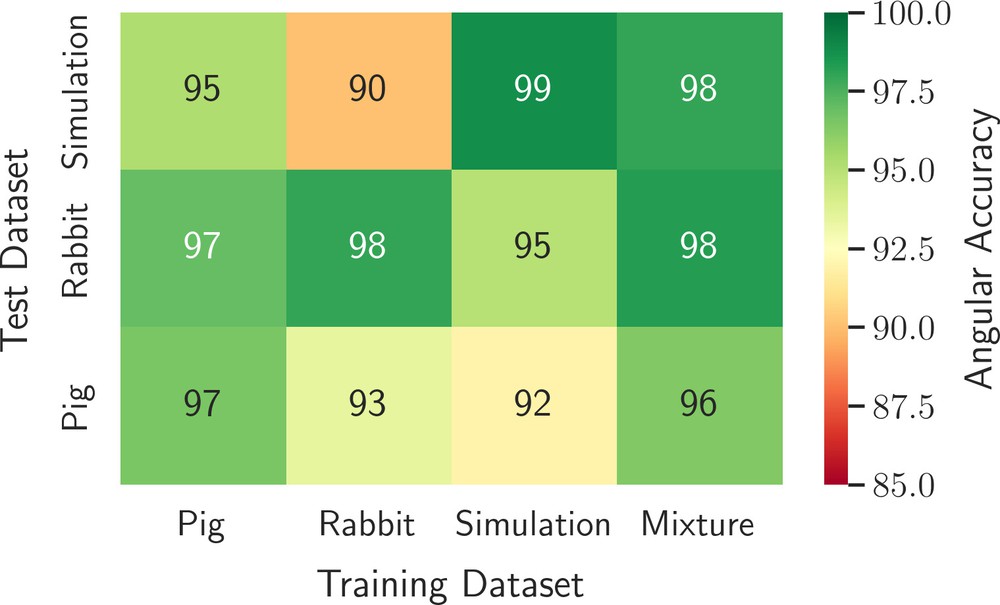}
  \caption{
  {
    Phase prediction accuracies for neural network models trained on either pig, rabbit or simulation data, or a mixture of all data, cf. Fig.~\ref{fig:FigureTrainingData} and Supplementary Video 2.
     Prediction across species or from simulation to experiment with models trained on either one species and applied to another species or on simulation data and applied to rabbit or pig optical mapping data.
    All models were applied to test data consisting of $5{,}000$ samples from experimental recordings or simulations, respectively.
    The prediction is most accurate when trained on the same data
    (Pig$\rightarrow$Pig $97\% \pm 8\%$;
    Rabbit$\rightarrow$Rabbit $98\% \pm 6\%$;
    Simulation$\rightarrow$Simulation $99\% \pm 4\%$).
    Nevertheless, the models appear to generalize as prediction across species is possible and achieves accuracies above $90 \%$ (Pig$\rightarrow$Rabbit $97 \% \pm 8\%$, Rabbit$\rightarrow$Pig $93 \% \pm 13\%$).
    The pig training data is more diverse than the rabbit training data (more hearts and different views), which yields higher accuracies when predicting from pig to rabbit than vice versa. 
    A model that was trained solely on simulation data can also be used to predict phase maps from experimental data (e.g. Simulation$\rightarrow$Rabbit $95\% \pm 10\%$).
  }
  }
\label{fig:FigureCrossTraining}
\end{figure}

{
We found that phase prediction models that were trained on pig optical mapping data can also be applied to rabbit optical mapping data and achieve equally high phase prediction accuracies on the data ($96.5\%\pm7.9\%$ vs. $97.0 \% \pm 7.5\%$), see Fig.~\ref{fig:FigureCrossTraining}. 
With such cross-species training, we observed higher accuracies when training from one species to another than vice versa (Rabbit$\rightarrow$Pig: $93.4\% \pm 12.2\%$ vs. Pig$\rightarrow$Rabbit: $97.0\% \pm 7.5\%$). This is presumably due to differences in the training data (more hearts, more diverse views in one species than the other).
Surprisingly, we found that even models that were solely trained with simulation data, as shown in Fig.~\ref{fig:FigureTrainingData}A), can be used to predict phase maps of VF optical mapping data and that these models achieve acceptable results, see lower row in Fig.~\ref{fig:FigureSimulationToExperiment}B) and Fig.~\ref{fig:FigureCrossTraining} (the simulation data was randomly masked with masks which were used with the experimental data, see Supplementary Video 2, all values outside the mask were set to $0$).
This demonstrates that the model can be applied to significantly different data than the data it was trained on. 
This also hints at the model generalizing and learning to associate phase maps with spatio-temporal dark-bright patterns in general rather than memorizing the particular wave dynamics.
Note that the simulation data only includes two-dimensional wave dynamics, whereas the experimental data corresponds to three-dimensional wave dynamics which are observed on the surface.
To our surprise, we found that models trained on simulation data without noise performed better on optical mapping data than when they were trained on simulation data with noise.
The network performed equally well across the different dynamical regimes in the simulated data, which includes episodes with both more laminar wave and more chaotic spiral wave dynamics with longer and shorter wavelengths, see Fig.~\ref{fig:FigureTrainingData}.
Lastly, Fig.~\ref{fig:FigureCrossTraining} shows that a neural network that was trained on a mixture of pig, rabbit and simulation data provides consistently high phase prediction accuracies of $96-98\%$ across all three datasets.
Taken together, these results demonstrate that the phase prediction neural network can be applied to a wide range of VF dynamics with various wave lengths and frequencies. 
Note that the rabbit data contains VF episodes with and without Cromakalim, which modulates the dynamics significantly. 
While it was not possible to create sufficiently large rabbit training datasets to determine the performance during cross-training (without Cromakalim $\rightarrow$ with Cromakalim or vice versa), we did not notice a significant change in accuracy when evaluating the performance of a general rabbit model on sub-data types (without Cromakalim vs. with Cromakalim). 
The analysis was performed with model M1.}

\quad
\subsection{Phase Singularity Prediction}

\begin{figure}[htb]
  \centering
  \includegraphics[width=14.0cm]{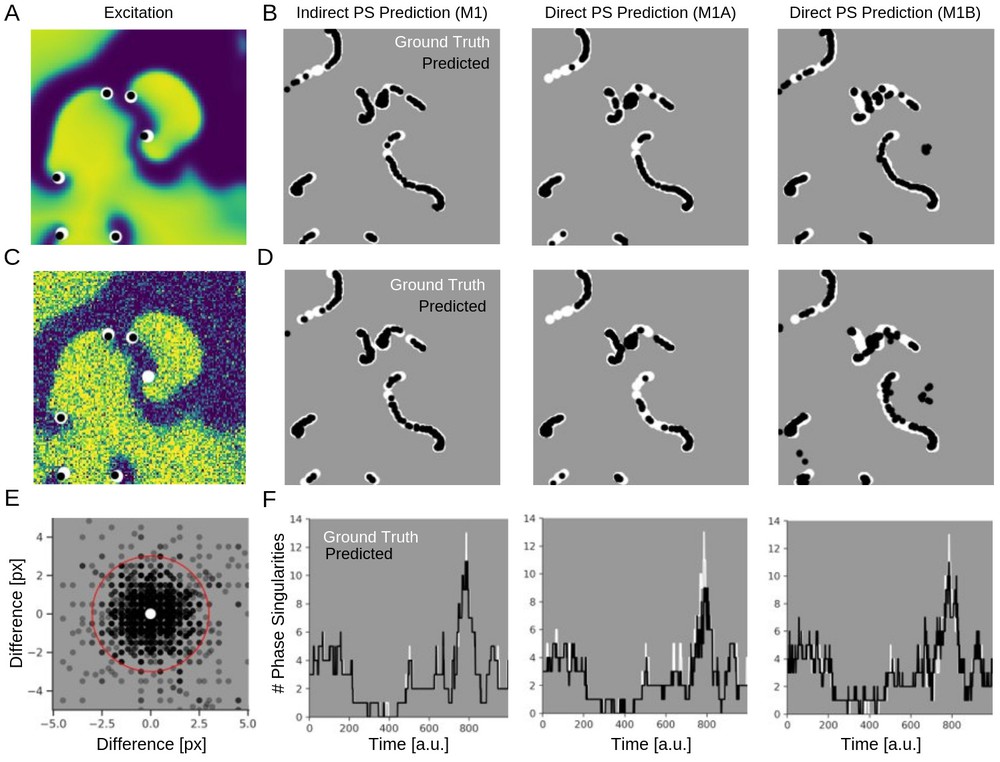}
  \caption{
  Phase singularities (PS) predicted for neural networks {M1 (indirect PS prediction, computing of PS from phase prediction), M1A (pixel-wise cross-entropy loss), and M1B (weighted Hausdorff distance loss)} from simulated maps of electrical excitation. 
  White: ground-truth or true PS. Black: predicted PS. {A quantitative evaluation of the predicted PS is shown in Table \ref{tab:pseval}}.
  A) Electrical spiral wave chaos without noise with PS superimposed indicating positions of spiral wave tips {for model M1A}.
  B) {Trajectories of ground truth (white) and predicted (black) PS without noise over 60 simulation time steps for the models M1, M1A, and M1B.}
  C) Electrical spiral wave chaos with noise ($\sigma=0.3$) with PS superimposed indicating positions of spiral wave tips.
  D) {Trajectories of ground truth and predicted PS with noise. Increase in false negative predictions with noise. Model M1B also produces false positive detections.}
  E) Spatial mismatch of predicted PS (black) and ground truth PS (white, center)  {for model M1A}. All predicted PS not within $3$ pixels (red circle) from true PS are false positives.
  F) Number of PS over time predicted {with models M1, M1A, M1B} from electrical spiral wave chaos with noise ($\sigma=0.3$).
  }
  \label{fig:FigureResults1}
\end{figure}

{
We found that the prediction of phase singularities (PS) from electrical excitation wave patterns was less accurate and less robust than predicting phase maps.
This was especially true with challenging data, such as optical mapping recordings, or noisy and sparsified simulation data.
Here, we compare three different neural network models M1, M1A, and M1B. 
Model M1 predicts PS indirectly by first predicting phase maps and subsequently calculating PS positions using the line integral technique.
Models M1A and M1B both predict PS directly, where M1A uses a pixel-wise loss function and M1B utilizes a distance-based loss function during training, see section \ref{sec:methods:networkarchitecture}.
Both models have different drawbacks: model M1A was better than M1B on simulation data both without and with noise, see Table~\ref{tab:pseval}, while M1B performs slightly better on optical mapping recordings than M1A, see Table~\ref{tab:PerformancePSPredictionRabbit} and Supplementary Video 3.
Model M1A is very conservative on challenging data, it occasionally produces false positives but mostly misses many true PS. 
Model M1B, on the other hand, is not as precise as M1A, and predicts more PS and produces more false detections.
Overall, the indirect PS prediction using model M1 shows the better performance than both direct methods with models M1A and M1B.}

\begin{table}[htb]
  \centering
\begin{tabular}{@{}lcccccccc@{}}\toprule
	Model     & \multicolumn{2}{c}{M1}   &\phantom{a}& \multicolumn{2}{c}{M1A}   &\phantom{a}& \multicolumn{2}{c}{M1B}   \\
	\cmidrule{2-3} \cmidrule{5-6} \cmidrule{8-9}
	Noise     & $\sigma=0$ & $\sigma=0.3$ &   & $\sigma=0$ & $\sigma=0.3$ &   & $\sigma=0$ & $\sigma=0.3$ \\\midrule
	Precision & 97.2\,\%   & 96.2\,\%     &   & 97.2\,\%   & 97.2\,\%     &   & 86.7\,\%   & 82.1\,\%     \\
	Recall    & 95.7\,\%   & 93.1\,\%     &   & 96.4\,\%   & 85.9\,\%     &   & 86.4\,\%   & 84.3\,\%     \\
	F-score   & 96.5\,\%   & 94.6\,\%     &   & 96.8\,\%   & 91.2\,\%     &   & 86.5\,\%   & 83.1\,\%     \\
	MAE       & 0.2        & 0.2          &   & 0.2        & 0.5          &   & 0.4        & 0.5          \\
	MAHD      & 2.3\,px    & 3.1\,px      &   & 2.0\,px    & 5.0\,px      &   & 4.0\,px    & 6.6\,px      \\
$\text{MAHD}^{\star}$ & 1.4\,px    & 1.8\,px      &   & 1.4\,px    & 3.1\,px      &   & 2.4\,px    & 3.6\,px\\\bottomrule
\end{tabular}
\caption{{
  Evaluation of phase singularity (PS) prediction on simulated electrical spiral wave chaos without ($\sigma=0$) and with noise ($\sigma=0.3$) for the different models M1, M1A, M1B. PS predictions were performed from 5 excitation frames and are shown in Fig.~\ref{fig:FigureResults1}.
  MAE is the mean absolute error of the number of predicted PS, MAHD is the mean average Hausdorff distance.
  The test dataset contains $5{,}000$ frames with $17{,}360$ PS in total, however $420$ frames contain no ground truth PS. 
  If the model predicts any PS for a frame which contains no ground truth PS -- or if no PS are predicted for a sample which does contain ground truth PS -- we assign a maximum average Hausdorff distance (181 px) for the computation of the MAHD. This skews the MAHD significantly. Accordingly, $\text{MAHD}^{\star}$ is the MAHD when we ignore these samples.
  }}
  \label{tab:pseval}
\end{table}

{Fig.~\ref{fig:FigureResults1} shows the PS predictions on simulated spiral wave chaos data. 
Panels A) and C) show predicted PS (black) and ground truth (white) PS superimposed onto the corresponding electrical excitation wave maps (PS were predicted with model M1A).
The maps demonstrate that both predicted and ground truth PS describe equally well the tips of spiral waves.
However, with noise, one of the six PS was not detected by the neural network (false negative detection).
Panels B) and D) show the trajectories of the predicted (black) and ground truth (white) PS over a short time span ($60$ simulation time steps), without and with noise ($\sigma = 0.3$), respectively, predicted indirectly with model M1 and directly with the models M1A and M1B.
The predictions were obtained from a short sequence of $N_t=5$ excitation frames, c.f. Fig.~\ref{fig:FigureSampling}A).
The trajectories co-align and demonstrate that PS are mostly predicted in locations where true PS are located. 
However, model M1B produces false positives even without noise.
Moreover, all PS prediction models miss a portion of ground truth PS, and we counted these mispredictions as false negatives.
Fig.~\ref{fig:FigureResults1}E) shows the spatial distribution of mismatches between predicted and ground truth PS for model M1A with noise $\sigma=0.3$, where the positions of the predicted PS are plotted relative to the position of the ground truth PS at the center.
All predicted PS which lie outside a radius of $3$ pixels (red circle) from the ground truth PS are counted as false positives. 
The sub-pixel resolution accuracy of PS is a result of our method: we calculated PS positions from a series of pixels in the PS prediction image, see section \ref{sec:methods:networkarchitecture}.
Table~\ref{tab:pseval} shows the evaluation of the PS prediction for all three models without and with noise in terms of precision, recall, F-score, MAE and MAHD on the test data consisting of $5{,}000$ samples with $17{,}360$ ground truth PS in total.
Without noise model M1A is slightly better or equal to the indirect model M1 (e.g. F-score of $96.8\,\%$ versus $96.5\,\%$), while model M1B is significantly worse in all measures (F-score $86.5\,\%$).
With noise however, the recall is significantly reduced for model M1A ($85.9\,\%$ versus $96.4\,\%$, F-score $91.2\,\%$), as the number of false negative predictions increases and the number of true positive predictions decreases (see Fig.~\ref{fig:FigureResults1}D).
The number of false negatives does not increase, however, and the precision stays the same without and with noise with model M1A.
This indicates that the model is rather conservative, insofar as when the difficulty for the model to predict PS locations increases it rather misses true PS instead of predicting false positives.
This can also be seen in Fig.~\ref{fig:FigureResults1}F), which shows the number of predicted (black) and ground truth (white) PS over time for $\sigma = 0.3$.
While the indirect PS predictions obtained with model M1 follow the ground truth PS closely, the direct PS predictions obtained with models M1A and M1B follow the trend overall but at times deviate considerably from the ground truth.
Model M1A consistently underestimates the number of PS, whereas model M1B both under- and overestimates PS.
Supplementary Video 5 shows the PS predictions with model M1A for different simulated electrical excitation wave patterns without and with noise as well as with sparsification.
The conservatism of model M1A is caused by its pixel-wise loss function, which does not account for the distance between predicted PS locations and true PS positions unless the pixels overlap.
The loss function is used during training to calculate an error value for every pixel of the predicted image of probable PS locations.
As the likelihood of a pixel containing a PS is very small, there is a class imbalance (number of pixels with versus without PS) for all pixel-wise loss functions and the network is biased towards not predicting a PS for challenging cases.
Model M1B, on the other hand, uses a loss function which is directly based on the distance between predicted and ground truth PS locations.
Table~\ref{tab:pseval} shows however, that model M1B is significantly less accurate for all measures than models M1 and M1A both with and without noise.
Panels B) and D) in Fig.~\ref{fig:FigureResults1} show that M1B predicts false positives both without and with noise.
In contrast, the indirect PS prediction with model M1 produces very few false positives, and the recall as well as the precision decrease only moderately with the addition of noise resulting in an F-score of $94.6\,\%$.
With regard to the direct PS prediction it is important to point out that the PS locations are determined by weighting multiple pixels, which surround the true PS and indicate probable PS positions, see Fig.~\ref{fig:FigureEncoding}C).
With model M1B this position estimation is especially problematic because a lot more pixels indicate the PS than with model M1A.
Accordingly, two nearby PS are often not sufficiently resolved in the prediction image and cannot be separated, which then produces a false positive detection between two true PS.
We tested different methods designed to extract individual PS positions as proposed by \citet{Ribera2019}, but did not observe an improvement of the PS prediction performance with model M1B.}

{
Table~\ref{tab:PerformancePSPredictionRabbit} and Supplementary Video 3 show PS predicted by the same models when trained and evaluated on rabbit optical mapping data, see also Fig.~\ref{fig:FigureTrainingData}. 
The indirect PS prediction with model M1 (F-score of $80.1\,\%$) is far more robust than and superior to the direct PS prediction with experimental data.
We found that these indirectly predicted PS matched the dynamics of the true PS computed from rabbit optical mapping data very well (see video).
The direct prediction model M1A performs poorly (F-score of $11.8\,\%$), as it misses most true PS (recall $11.8\,\%$). 
However, it appears to predict some false positives mainly at the medium boundaries (see video).
Model M1B achieves a significantly better F-score than model M1A of $42.5\,\%$ on the optical mapping data, as it does not suffer from the conservatism exhibited by model M1A.
However, overall, the performance of model M1B is still poor on optical mapping data.
}

\begin{table}
  \centering
\begin{tabular}{lccccc}
  \toprule
  Model &  Precision &  Recall &  F-Score &  MAE & MAHD \\
  \midrule
    M1 &       82.9\,\% &    77.5\,\% &    80.1\,\% &  0.7 &  11.6 \\
    M1A &      41.9\,\% &    11.8\,\% &    18.4\,\% &  2.2 & 113.6 \\
    M1B &      40.4\,\% &    44.9\,\% &    42.5\,\% &  1.4 &  24.8 \\
  \bottomrule
\end{tabular}  
\caption{{
  PS prediction with models M1 (indirect from phase), M1A (pixel-wise loss), and M1B (distance-based loss) when trained and evaluated on rabbit optical mapping data using a radius of $r=6\,\text{px}$ for computation of precision, recall and F-score. Supplementary Video 3 shows the predicted PS for all three models. The indirect PS estimation (M1) is far more accurate than the direct PS prediction (M1A and M1B).
}}
\label{tab:PerformancePSPredictionRabbit}
\end{table}

\begin{figure*}[htb]
  \centering
  \includegraphics[clip, trim=0.0cm 0.0cm 0.0cm 0.0cm, width=16cm]{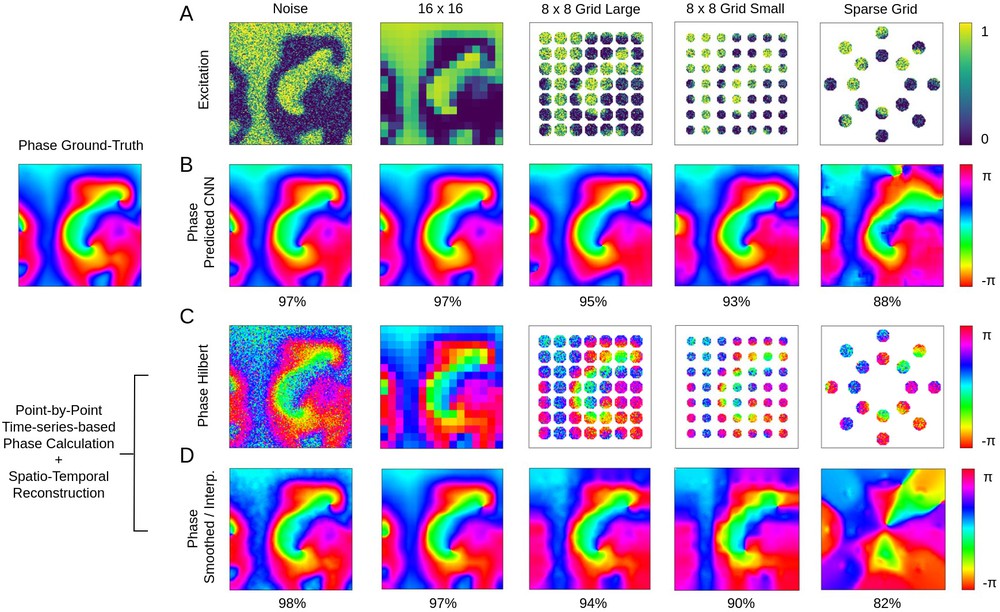}
  \caption{
  {
  Deep learning-based prediction of phase maps from noisy and/or sparse electrical excitation wave patterns. 
  Left: Corresponding ground-truth phase map $\phi(x,y)$ calculated from original electrical excitation wave pattern (at $t=415$) without noise or sparsification via the Hilbert transform, as shown in Fig.~\ref{fig:Figure01Hilbert}A,B).
  A) Excitation wave patterns with noise ($\sigma=0.3$), low resolution (no-noise excitation pattern down-sampled with averaging to $16{\times} 16$ pixel then up-sampled to $128{\times}128$ pixel), $8{\times}8$ grid of large round electrodes or fiber optics ($15$ pixel diameter, $43\%$ coverage, $\sigma=0.3$), $8{\times}8$ grid with small round electrodes or fiber optics ($11$ pixel diameter, $21\%$ coverage, $\sigma=0.3$) and a sparse star-shaped / ring-shaped grid of large round electrodes or fiber optics ($15$ pixel diameter, $16\%$ coverage, $\sigma=0.3$).
  B) Corresponding predicted phase maps $\hat{\phi}(x,y)$ with $96.8\% \pm 3.2 \%$, $96.8\% \pm 3.4 \%$, $94.6\% \pm 7.4 \%$, $93.1\% \pm 9.1\%$ and $87.8\% \pm 15.4 \%$ angular accuracies from left to right, respectively. Except with the sparse grid, the predicted phase maps $\hat{\phi}$ are hard to distinguish from the true phase map $\phi$. The data was not seen by the network during training. Phase maps $\hat{\phi}(x,y,t_p)$ were predicted from a short spatio-temporal sequence of $5$ electrical excitation wave maps $V(x,y,t=t_1,t_2,t_3,t_4,t_5)$.
  C) Phase maps of the noisy, low resolution and sparse excitation wave patterns calculated via the Hilbert transform.
  D) Smoothed and/or interpolated versions of the phase maps shown in C) with $97.7\% \pm 2.9 \%$, $96.9\% \pm 5.2 \%$, $93.9\% \pm 10.3 \%$, $90.1 \pm 13.1 \%$ and $81.9 \% \pm 21.0 \%$ angular accuracies from left to right, respectively. 
  Kernel-based phase smoothing and interpolation methods described in section \ref{sec:methods:interpolation}. Note that the phase maps were calculated from video data and not from just $5$ snapshots like in B).}
  }
  \label{fig:FigureResults2Phase}
\end{figure*}

\quad
\subsection{Prediction of Phase Maps from Noisy, Low-Resolution or Sparse Excitation Wave Maps}

\begin{figure}[htb]
  \centering
  \includegraphics[clip, trim=0.0cm 0.0cm 0.0cm 0.0cm, width=8.4cm]{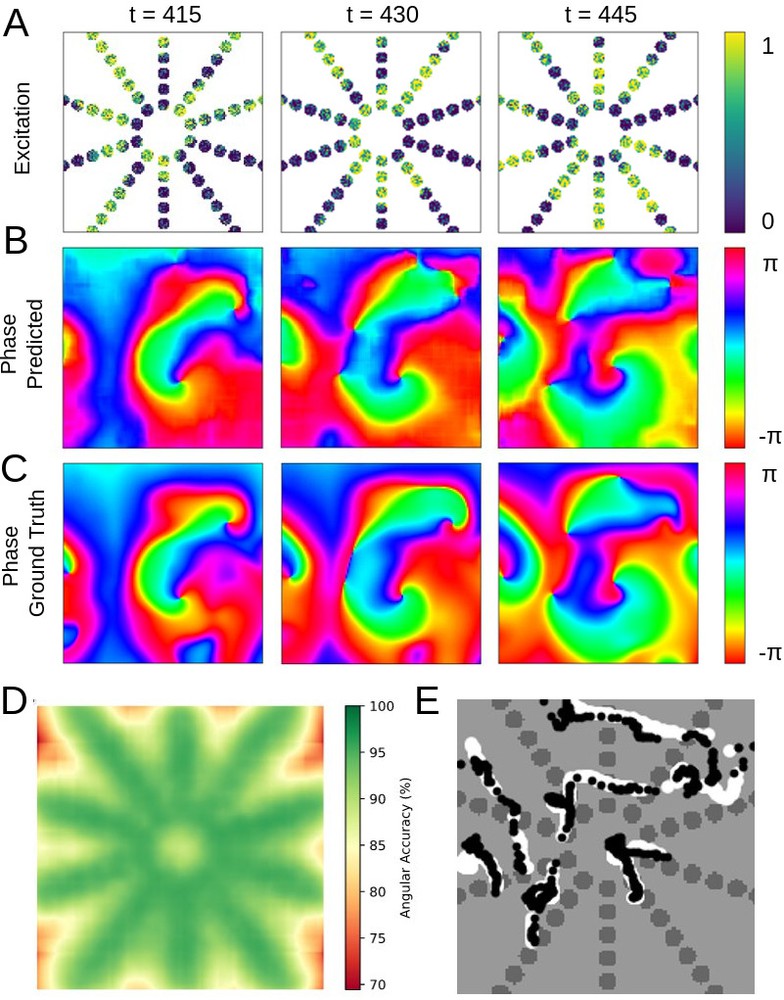}
  \caption{
  Deep learning-based prediction of phase maps and rotor cores or phase singularities (PS) from sparse electrical excitation wave pattern mimicking multi-electrode catheter or optical fiber recordings.
  A) Sparse excitation wave pattern with noise ($\sigma = 0.3$, 17\% coverage). 
  B) Phase map $\hat{\phi}(x,y)$ predicted by neural network analyzing data in A). %
  C) Ground-truth phase map $\phi(x,y)$ obtained with complete, non-sparse, non-noisy data. 
  {
  D) Spatially resolved angular accuracy (temporal average in each pixel) shows that accuracy decreases between electrodes.
  E) Trajectories of ground truth PS (white) and predicted PS (black) using indirect prediction with model M1 (shown over $100$ simulation time steps), see also Supplementary Video 8.
  }
  }
  \label{fig:FigureResults3Phase}
\end{figure}

{The phase prediction neural network can predict phase maps even from very noisy, low-resolution and/or very sparse electrical excitation wave maps.
Figs.~\ref{fig:FigureResults2Phase}-\ref{fig:FigureResults3Phase} and Supplementary Videos 4 and 6 show phase predictions obtained with model M1 with various simulated noisy, low-resolution or sparse excitation wave patterns, which are very generic simulations of imaging scenarios with low-resolution or low signal-to-noise sensors, multi-electrode arrays or (catheter) mapping electrodes, fiber optics or other similar sensors.
Panel A) in Fig.~\ref{fig:FigureResults2Phase} shows exemplary snapshots of the excitation videos that were analyzed: 1) a noisy ($\sigma = 0.3$) excitation pattern with $128{\times}128$ pixels resolution, 2) a low-resolution version of the same pattern that was derived by down-sampling the original non-noisy excitation pattern to $16{\times}16$ pixels resolution and then up-sampling the pattern without interpolation to $128{\times}128$ pixels resolution, 
3) a $8{\times}8$ grid of large round electrodes or fiber optics $16$ pixels apart with a diameter of $15$ pixels each, the grid covering $43\%$ of the area (with noise $\sigma=0.3$), 
4) a $8{\times}8$ grid with small round electrodes or fiber optics $16$ pixels apart with a diameter of $11$ pixels each, the grid covering $21\%$ of the area (with noise $\sigma=0.3$),
and 5) a sparse star-shaped / ring-shaped grid of large round electrodes or fiber optics with a diameter of $15$ pixels each, the grid covering $16\%$ of the area (with noise $\sigma=0.3$).
Panel B) in Fig.~\ref{fig:FigureResults2Phase} shows the corresponding predicted phase maps $\hat{\phi}$ predicted using the neural network model M1.
The predicted phase maps $\hat{\phi}$ are visually nearly indistinguishable from the ground truth phase map $\phi$ shown as a reference on the left.
The phase maps were predicted with angular accuracies of $96.8 \% \pm 3.2 \%$, $96.8 \% \pm 3.4 \%$, $94.6 \% \pm 7.4 \%$, $93.1 \% \pm 9.1\%$ and $87.8 \% \pm 15.4 \%$ from left to right, respectively.
The maps illustrate that the deep-learning-based phase prediction can suppress noise, enhance spatial resolution, and interpolate missing data and recover phase maps even when it only sees a fraction of the electrical data, as shown in the last example and in Fig.~\ref{fig:FigureResults3Phase}.} 

{The ground truth phase map $\phi$ in Fig.~\ref{fig:FigureResults2Phase} was computed from the original electrical excitation wave pattern without noise ($\sigma = 0.0$) using the Hilbert transform, computing in each pixel $(x,y)$ individually a phase signal from time-series data $V(t)_{x,y} \rightarrow \phi(t)_{x,y}$ as shown in Fig.~\ref{fig:Figure01Hilbert}.
The phase maps shown in panel C) in Fig.~\ref{fig:FigureResults2Phase} were equally computed pixel-by-pixel using the Hilbert transform, but were computed directly from the noisy, low-resolution or sparse electrical excitation data shown in A). The phase maps accordingly include the same features, e.g. they include noise or remain sparse.
The phase maps shown in panel D) were reconstructed from the noisy, low-resolution or sparse phase maps in C) using spatio-temporal inpainting and smoothing techniques, as described in section \ref{sec:methods:interpolation}.
They serve as a reference and allow the comparison of the deep learning with another interpolation method.
While the reconstructed phase maps in D) also provide sufficiently accurate reconstructions with noise, low-resolution and low sparsity, with further increasing sparsity the reference method fails to produce accurate results and is outperformed by the deep learning-based phase prediction.
Note that, even though the accuracy of both approaches are equally or comparably high, the reconstructed phase maps in D) contain noise or distortions, while the deep learning-based approach in B) produces consistently very smooth phase maps.}

{The neural network's ability to interpolate and reconstruct phase maps allows the tracking of PS between sensors even when they are relatively far apart, see Fig.~\ref{fig:FigureResults3Phase} and Supplementary Video 8.
Panels A-C) in Fig.~\ref{fig:FigureResults3Phase} show a sparse, star-shaped electrode/sensor configuration measuring an excitation wave pattern, and the resulting predicted and ground truth phase maps with this configuration, respectively.
The predicted phase maps resolve the rotor dynamics very well, particularly towards the center where the electrode density is higher (average angular accuracy for entire field of view: $\sim91\%$).
The average angular accuracy (temporal average in each pixel) resolved in space in panel D) indicates that the phase prediction accuracy remains sufficiently high between the electrodes towards the center.
Accordingly, panel E) shows how the predicted PS (black) represent the ground truth PS (white) sufficiently well and follow their trajectories between and across electrodes (shown over $100$ simulation time steps).
The total area of the sensor/electrodes covers only $17\%$ of the entire 2D simulation domain.}

\quad
\subsection{Extreme Sparsity and Noise}

\begin{figure}[htb]
  \centering
  \includegraphics[clip, width=0.6\textwidth]{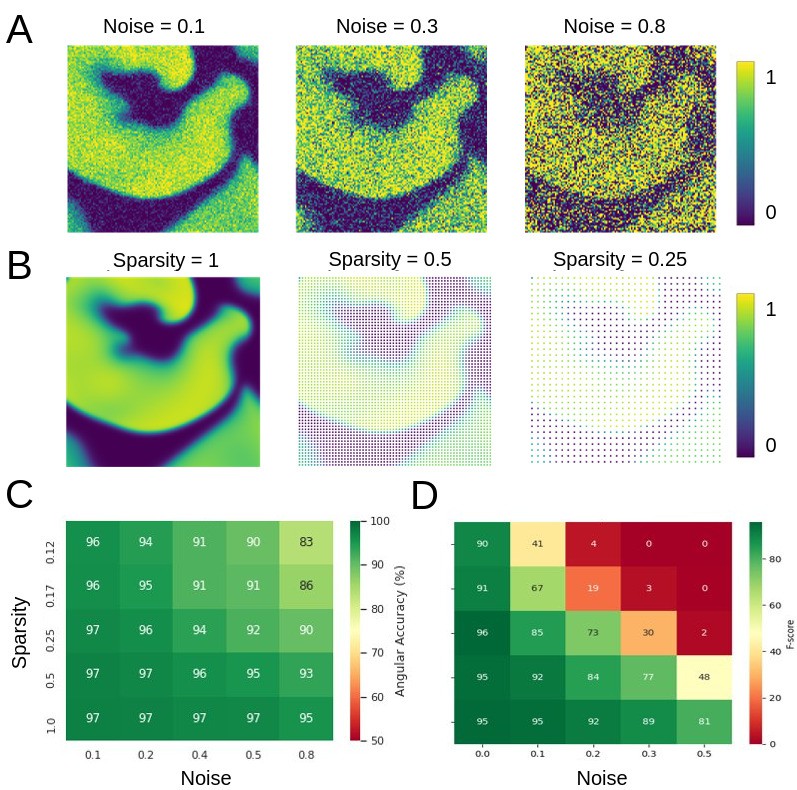}
  \caption{
  Prediction accuracies with noisy and sparse data.
  A) Noisy simulated electrical excitation wave patterns with noise levels of $\sigma = 0.1, 0.3, 0.8$.
  {B) Sparse simulated electrical excitation wave patterns with sparsity levels of $\xi=1, 0.5, 0.25$. With $\xi=0.25$ all pixels except every 4th pixel are set to $0$ (no signal).
  C) High phase prediction accuracies across broad range of noise and sparsity levels with model M1 (here shown with $N_t=5$ sampled images, which are $\tau=5$ simulation time steps apart, c.f. Fig.~\ref{fig:FigureSampling}).}
  D) PS prediction accuracy obtained with model M1A is more sensitive to noise and sparsity. The direct PS prediction fails when data is both very noisy and/or sparse.
  }
  \label{fig:FigureNoiseVsSparsity}
\end{figure}

{
The data shown in Fig.~\ref{fig:FigureNoiseVsSparsity} characterizes  the phase and PS prediction performance with extreme noise and sparsity in more detail.
Panel A) shows maps with simulated electrical excitation wave patterns with noise levels of $\sigma = 0.1, 0.3, 0.8$, and panel B) shows the same simulated electrical excitation wave patterns without noise but sparsified with sparsity levels of $\xi= 1.0, 0.5, 0.25$.
The excitation images were sparsified by setting all pixels except every n-th pixel in $x$- and $y$-direction to $0$ (no signal). 
Accordingly, a sparsity level of $\xi=0.25$ corresponds to setting every pixel but every 4th pixel to $0$, for instance.
Panels C) and D) show the prediction accuracies obtained with different combinations of noise and sparsity for the phase prediction with model M1 (angular accuracy) and for the PS prediction with model M1A (F-score), respectively.
The individual prediction accuracies were obtained when training was performed with each specific combination of $\sigma$ and $\xi$.
The map in C) shows that the phase prediction with model M1 is highly accurate and remains above $90\%$ angular accuracy over a wide range of noise and sparsity levels.
In particular, with non-sparse data ($\xi = 1.0$), the angular accuracy stays above $95\%$ with noise levels of up to $\sigma = 0.8$, which corresponds to the noise level shown on the right in panel A).
With $\xi = 0.125$ sparsification, the information in the image is reduced to $16{\times}16 = 256$ non-zero pixels instead of $128{\times}128 = 16,384$ pixels. Therefore, the neural network can analyze only less than $2\%$ of the image. Despite this reduction, the network provides accuracies of $94-97\%$ with noise levels of $\sigma=0.1-0.2$ (and at least $90\%$ with noise levels of up to $\sigma=0.5$).
While the phase prediction is accurate over a broad range of noise and sparsity levels, panel D) shows that the direct prediction of PS using model M1A is less accurate and robust against noise or sparsity.
The F-score stays above $90\,\%$ only at low noise or sparsity levels and deteriorates quickly when both increase (e.g. sparsity $\xi=0.25$ and noise $\sigma=0.3$). 
The F-score even drops entirely to $0\,\%$ in extremely noisy and sparse regimes.
The systematic analysis in panel D) confirms the impression given in Fig.~\ref{fig:FigureResults1}C,D) that the predicted PS trajectories frequently contain false detections when predicted directly.
The angular accuracies and F-scores were computed over the testing dataset with $5{,}000$ frames.
The data shows that even though noise and sparsity impair the phase prediction accuracy, overall the phase predictions remain, in contrast to the PS prediction, robust and sufficiently accurate in the presence of strong noise and with extreme sparsity. 
Autoencoder neural networks have excellent denoising capabilities and are very effective at interpolating image data \citep{Vincent2008,Gondara2016}, and this property can be observed at work in Figs. \ref{fig:FigureExperiments}, \ref{fig:FigureResults2Phase} and \ref{fig:FigureNoiseVsSparsity}.}

\quad
\subsection{Spatio-Temporal Sampling over Spiral Wave's Period increases Prediction Accuracy}\label{sec:results:spatiotemporal}

\begin{figure}[htb]
  \centering
  \includegraphics[clip, width=0.5\textwidth]{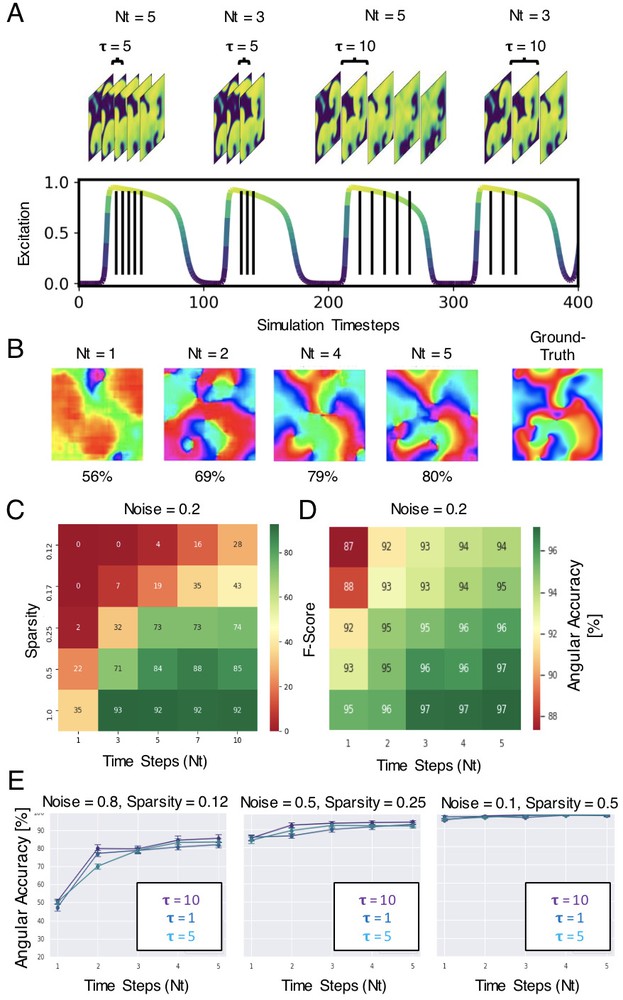}
  \caption{
  Spatio-temporal sampling of excitation wave dynamics. 
  A) Schematic drawing illustrating number of sampled frames $N_t$ and sampling distance $\tau$ between the frames with $N_t=3$, $N_t=5$ and $\tau=5$, $\tau=10$, respectively, shown relative to action potential duration (action potential duration $\sim 75-80$ time steps, cycle length or period $\sim 100$ time steps).
  {B) Analyzing longer temporal sequences with $N_t=5$ frames increases prediction accuracy, here shown for very noisy ($\sigma=0.8$) and sparse data ($\xi=0.125$ or $16 \stimes 16$ non-zero data points), c.f. Fig.~\ref{fig:FigureNoiseVsSparsity}A-C). In this example, the sampled excitation snapshots are $\tau=5$ simulation time steps apart.}
  C) PS prediction accuracy (F-score) over number of samples $N_t$ and sparsity $\xi$. More samples increase accuracy (here shown for noise $\sigma=0.2$).
  D) Phase prediction accuracy (angular accuracy) over number of samples $N_t$ and sparsity $\xi$. More samples increase accuracy (here shown for noise $\sigma=0.2$).
  E) Phase prediction accuracy is not (significantly) affected by variation of sampling distance $\tau$ (all curves overlap, here shown with $\tau=1,5,10$).
  }
  \label{fig:FigureSampling}
\end{figure}

{The neural network does not require very much information to be able to predict phase maps or phase singularities (PS). 
A short sequence of $N_t=5$-$10$ excitation wave patterns is sufficient in most situations to make accurate predictions, even with extreme noise and/or sparsity, as shown in Fig.~\ref{fig:FigureNoiseVsSparsity}. 
The results in Figs.~\ref{fig:FigureExperiments}-\ref{fig:FigureNoiseVsSparsity} were obtained with either $N_t=5$ or $N_t=10$ excitation frames with simulated or experimental data, respectively.
The number of sampled frames $N_t$ and the sampling distance $\tau$, which corresponds to the temporal offset between the samples, see sketch in Fig.~\ref{fig:FigureSampling}A), are the two main parameters determining the phase and PS prediction accuracy.}

{Regarding the number of sampled frames $N_t$, we found that the predictions become more accurate when sampling the activity with more frames, but the accuracy does not improve significantly further with more than $5-10$ frames. 
Analyzing a short spatio-temporal sequence ($N_t=4,5,...,10$) rather than just a single, static ($N_t=1$) excitation wave pattern or a few ($N_t=2,3$) excitation wave patterns does not only increase the accuracy, but also improves the prediction robustness and ensures that the neural network is able to make predictions at all in difficult environments with high noise or sparsity, see Fig.~\ref{fig:FigureSampling}B-D).
Fig.~\ref{fig:FigureSampling}B) shows how the prediction fails entirely if the network only analyzes $1$ frame, but becomes progressively better with each frame and finally succeeds to produce satisfactory phase predictions ($80\%$) when it analyzes a short sequence of $N_t=2,4,5$ frames. 
In this example, the phase map was predicted from a very noisy ($\sigma=0.8$) and very sparse ($\xi=0.125$) excitation pattern.
The data demonstrates that the neural network is able to compensate information that is lacking in space with additional information it retrieves over time.
The multi-frame analysis can also slightly improve the neural network's prediction accuracy when it already achieves high accuracies in less extreme conditions.
Fig.~\ref{fig:FigureSampling}C) shows that the F-score increases from $22\%$ to about $85\%$ when using $N_t=1,3,5,7,10$ frames for the direct PS prediction with model M1A. 
Fig.~\ref{fig:FigureSampling}D) shows that the angular accuracy increases slightly from $93\%$ to $97\%$ when using $N_t=1,2,3,4,5$ frames for the phase prediction with model M1 with low sparsity ($\xi=0.5$) and low noise ($\sigma = 0.2$).
The PS prediction benefits more from the multi-frame analysis as it is more sensitive to noise and sparsity.}

{Regarding the sampling distance $\tau$, we made the following observations: 
1) With experimental data, we were able to mix the rabbit, pig and simulation data and even though $\tau$ was not perfectly adjusted to all of the different dominant frequencies of the wave dynamics or imaging speeds, the network was able to produce accurate predictions across all data, see Fig.~\ref{fig:FigureCrossTraining}.
We chose a sampling distance of $\tau=12\,\text{ms}$ for both the rabbit and pig data, resulting in an effective framerate of $83\,\text{fps}$.
With $N_t=10$ the series of sampled frames covered $75$-$140\%$ of the cycle length or dominant period of the VF dynamics (about $90\,\text{ms}$-$170\,\text{ms}$).
The phase prediction failed with the experimental data when we used shorter sampling times $T_{\tau} = \tau \cdot N_t$, which covered only a smaller fraction of the cycle length (e.g. $15\%$).
2) With the simulation data shown in Fig.~\ref{fig:FigureTrainingData}A), the sampling distance $\tau$ did not affect the phase prediction accuracy at all, and we achieved high accuracies even with short sampling times $T_{\tau}$ (e.g. $5\%$ with $N_t=5$ and $\tau = 1$). 
Fig.~\ref{fig:FigureSampling}E) shows that the angular accuracy does not change significantly when the sampling distance is varied ($\tau = 1$, $\tau = 5$ or $\tau = 10$ with $N_t=5$ frames, shown for $3$ different noise and sparsity levels).
With these parameters, the dynamics are sampled over $5$, $25$ or $50$ simulation time steps, which corresponds to about $5\%$, $25\%$ or $50\%$ of the average cycle length or dominant period of the spiral wave dynamics of about $100$ simulation time steps, respectively.}

\quad
\subsection{Training Data Diversity increases Robustness against varying Imaging Parameters}

\begin{figure}[htb]
  \centering
  \includegraphics[clip, width=0.7\textwidth]{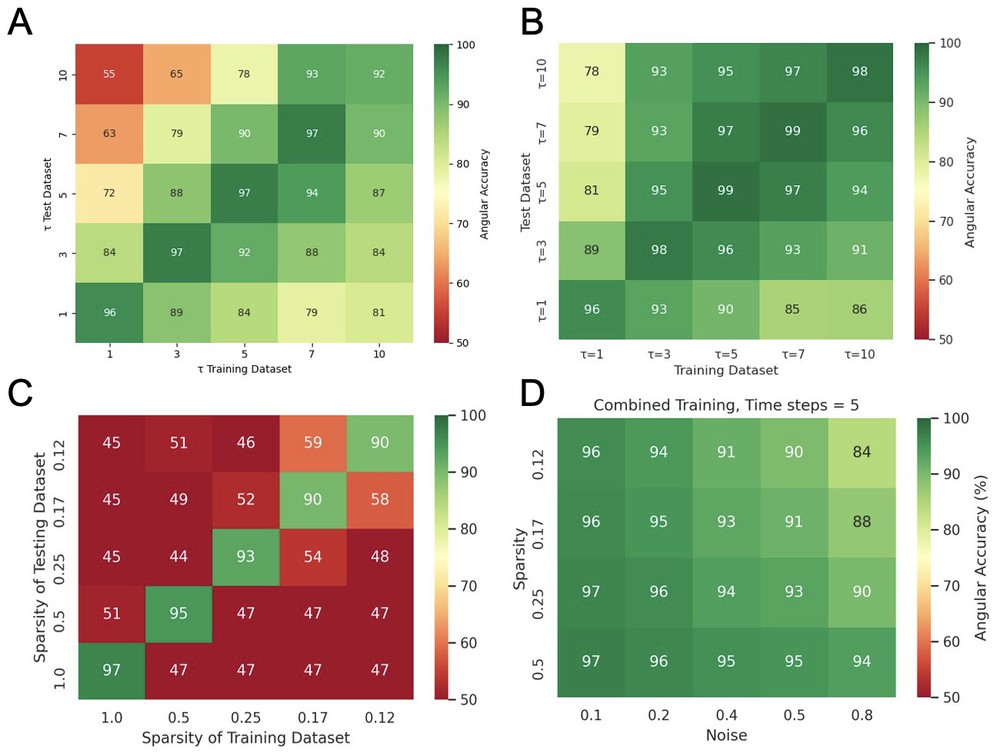}
  \caption{
  {Diverse and augmented training data increases robustness of deep-learning-based phase prediction. 
  A) The phase prediction accuracy decreases if a model that was trained with a specific sampling distance $\tau$ is applied to data that was sampled with a different sampling distance.
  B) Data augmentation (randomly masking the data as shown in the right panel in Supplementary Video 2) minimizes the effect in A).
  C) Phase prediction fails when neural network is trained on data with one specific sparsity, see also Fig.~\ref{fig:FigureNoiseVsSparsity}B), and is then applied to data with different sparsity (off diagonal).
  D) Phase prediction stays accurate across all noise levels and sparsities when training data also includes all noise and sparsity levels. Note that, by contrast, in Fig.~\ref{fig:FigureNoiseVsSparsity}C) the accuracy map was created by training separate models individually with each noise and sparsity combination.
  All results obtained with simulation data and neural network model M1.}
  }
  \label{fig:FigureGeneralization}
\end{figure}

{Training the neural network with more diverse data broadens the distribution of data it can analyze and will prevent eventual overfitting to a particular feature in a dataset.
The network's insensitivity to the sampling distance $\tau$ with the simulation data, as discussed in section \ref{sec:results:spatiotemporal} and shown in Fig.~\ref{fig:FigureSampling}E), is an indication for overfitting when training and predicting solely on simulation data, because the same model trained with and applied to optical mapping data is unable to produce correct predictions with short $\tau$. The different behavior with simulation and experimental data suggests that the network specializes with the simulation data in memorizing the dynamics based on instantaneous features (moving wavefronts etc.). However, this approach fails with experimental data, in which case it only succeeds if it is provided information that was sampled over a significant portion or the entire period of the reentry pattern.
Interestingly, we also made the following observation: panel C) in Fig.~\ref{fig:FigureGeneralization} shows that the phase prediction accuracy drops if training was performed on the simulation data with just one specific sampling distance $\tau_{train}$ and the network is then applied to data that was sampled with a different sampling distance $\tau \neq \tau_{train}$. Importantly, the analysis was performed on the simulation data without data augmentation, as shown in Fig.~\ref{fig:FigureTrainingData}A) (with $\sigma=0$, $\xi=1$). 
However, panel D) shows that if the same simulation data is augmented with the masks shown in Fig.~\ref{fig:FigureTrainingData}B), see also Supplementary Video 2, then the network performs better and achieves higher accuracies at other sampling distances $\tau$ even though it was not trained on these $\tau$ values. For instance, if the network was trained with $\tau_{train}=5$ and achieves an accuracy of $99\%$ at $\tau=5$, it still achieves an accuracy of $96-97\%$ with $\tau=3$ or $\tau=7$ just because the input data was augmented and includes other features (arbitrary masked regions) than just the wave dynamics on a square simulation domain. These findings are consistent with the finding that a single $\tau$ could be used with a mix of experimental and simulated data, as described in section \ref{sec:crosstraining}.}

{Similarly, panel C) in Fig.~\ref{fig:FigureGeneralization} shows that if the neural network is trained solely with a particular sparsification, for instance with $\xi=0.25$, then it excels at performing predictions with $\xi=0.25$, but fails with different sparsifications. 
Accordingly, the phase prediction only succeeds when the sparsity of the testing dataset matches the sparsity during training (along the diagonal), and fails when the sparsities in the training and testing datasets are different (off the diagonal).
However, this issue can be resolved by training the network with data that includes all sparsifications (here $\xi=1.0,0.5,0.25,0.17,0.125$).
Panel D) in Fig.~\ref{fig:FigureGeneralization} shows that the same neural network can be applied to arbitrary noise and sparsification levels and will consistently yield phase prediction accuracies above $90\%$ when the training was performed with data that contained all noise and sparsification levels. 
By contrast, in Fig.~\ref{fig:FigureNoiseVsSparsity}C) the training was performed individually with each specific combination of noise and sparsification. 
The broader training in Fig.~\ref{fig:FigureGeneralization}D) makes the network more robust and yields just as high phase prediction accuracies as with each individual specialized training in Fig.~\ref{fig:FigureNoiseVsSparsity}C).}

The anecdotal findings in Figs.~\ref{fig:FigureCrossTraining} and \ref{fig:FigureGeneralization} are representative of a very general property of neural networks and data driven approaches, and similar observations would be made with other parameters, such as noise, blurring or arbitrary sparsification patterns and we made very similar observations in a previous study \citep{Christoph2020} with an architecturally very similar neural network.

\quad
\subsection{Predicting Future Phase Maps or PS Positions}

\begin{figure}[htbp]
  \centering
  \includegraphics[clip, width=0.83\textwidth]{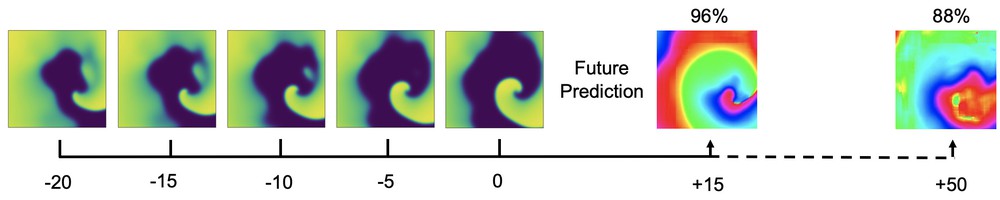}
  \caption{
  {Prediction of phase maps in future time steps with simulated electrical spiral wave chaos. Phase prediction accuracies of $98\%$ and $88\%$ predicting $15$ and $50$ simulation time steps into the future, respectively, analyzing $N_t=5$ excitation wave frames at $t=0,$$-5$,$-10$,$-15$,$-20$. The average rotational period of the spiral waves are about $100$ simulation time steps.}
  }
  \label{fig:FigureFuture}
\end{figure}

{It is possible to predict phase maps and PS positions in future time steps, but only within the immediate future. 
Fig.~\ref{fig:FigureFuture} shows predictions of phase maps with simulated spiral wave chaos, which the neural network M1 predicted $15$ and $50$ simulation time steps into the future. 
The network analyzed $N_t=5$ excitation wave frames at $t=0$,$-5$,$-10$,$-15$,$-20$ to make each phase prediction and achieved $97.6\% \pm 2.5 \%$, $96.0\% \pm 3.4 \%$, $95.5 \% \pm 4.1 \%$ and $87.7\% \pm 17.6 \%$ angular prediction accuracy $5$, $10$, $15$ and $50$ simulation time steps into the future, respectively. As the average cycle length or rotational period of the activity is about $100$ simulation time steps, this corresponds to about half a rotation within which the prediction yet achieves satisfactory accuracies and about 1/5 of a rotation within which the prediction achieves very good accuracies.}

\section{Discussion}

We demonstrate that deep neural networks can be used to compute phase maps and locate the position of phase singularities (PS) when analyzing cardiac excitation wave dynamics. 
{PS can be predicted by deep neural networks either directly from excitation wave patterns or indirectly by predicting first phase maps from the excitation wave patterns and then calculating PS in the predicted phase maps using classical techniques (e.g. the circular line integral method, shown in Fig.~\ref{fig:Figure01Hilbert}C)). 
This latter step is possible because the predicted phase maps are smooth. 
We found that the direct PS prediction was less robust than the prediction of phase maps, particularly with challenging data, and, accordingly, we only succeeded to predict PS positions reliably in experimental data with the indirect method.}
Predictions of phase maps and PS can be performed almost instantaneously from a short temporal sequence consisting of $1$-$10$ snapshots of cardiac excitation waves.
We successfully applied this {deep learning-based} rotor localization and phase mapping technique to both simulated and {ex-vivo optical mapping data} of ventricular fibrillation (VF), {and we expect that the technique can also be applied to catheter mapping data of cardiac arrhythmias in clinical patients.}

One of the most critical issues with neural networks is their ability 'to generalize'. 
Neural networks are known to perform very well when applied to data that is very similar to, or 'within the distribution', of the training data, but their accuracy and robustness can quickly deteriorate when applied to other, less similar 'out-of-distribution' data.
{Our results demonstrate that our deep learning-based phase mapping algorithm can be developed in one species and then applied to another species. 
We even show that the phase mapping algorithm can be developed with synthetic data generated in computer simulations and then applied to experimental data. This latter observation is particularly noteworthy in that the simulation data used to train the network was 2D, whereas the experimental data to which it was applied were surface observations of 3D dynamics.
These findings suggests that the algorithm is able to learn the relevant correlation between patterns in a specific distribution of data, 
and then extrapolate this mapping to differently distributed data that is well outside of the training distribution.
From our results it appears that the deep learning algorithm learns to associate phase patterns with a broad class of excitable spatio-temporal activity, and understands the more generalized phase mapping problem, independent of physiological parameters or species-dependent wave dynamics.} 

{Based on these findings, we anticipate that it will be possible to develop a similar deep learning-based phase mapping approach for clinical mapping of arrhythmias in human patients.
Neural networks can in principle analyze any data, and they will likely be able to predict phase maps from extracellular field potential or electrogram measurements, just as they are able to predict phase maps from optical measurements of the cellular transmembrane potential. 
Because neural networks excel at detecting hidden patterns in data, 'ignoring' noise, interpolating missing data, and enhancing spatial resolution, all of which they can do simultaneously, 
they are ideally suited for the analysis of catheter mapping data of atrial fibrillation.
The application of such a deep learning-based algorithm would not only be restricted just to phase mapping, but could in principle also be extended to map any other characterizing feature of arrhythmias (e.g. activation or conduction velocity maps).
As our results indicate, neural networks would be able to integrate sparse data acquired with multi-electrode basket catheters, given that they are trained with adequate high-resolution imaging data, which could be generated {\it ex vivo} or in computer simulations.
Ultimately, deep-learning has great potential to alleviate some of the shortcomings of catheter mapping, which are largely associated with limited spatial resolution and interpolation artifacts \citep{Martinez-Mateu2018,VanNieuwenhuyse2021}, that in turn can lead to misrepresentations of rotor dynamics and fibrillatory wave patterns during atrial fibrillation.
}

{Other advantages of our technique are i) that it can compute phase maps and PS in real-time with data that was acquired over a brief interval and ii) that it can obviate pre-processing of the raw data (e.g. spatio-temporal smoothing, outlier removal, etc.). 
The predictions do not require the collection of long time-series data, can be performed within one rotational period of the wave dynamics, see Figs.~\ref{fig:FigureSampling}A), and can be calculated in real-time at $500$-$1000\,\text{fps}$ using GPU hardware (at $128{\times}128$ pixels resolution). 
Predictions can furthermore be performed into the immediate future, enabling predictions of PS positions within about the next 1/4 rotation of reentrant wave dynamics, see Fig.~\ref{fig:FigureFuture}.
The latter aspect ii) makes the technique ideal for the processing of very noisy video data or data containing other artifacts, such as motion artifacts. 
This could also make it an attractive phase mapping approach in other fields beyond cardiovascular research, for instance, when studying the dynamics of excitation waves and topological defects in other biological systems \citep{Huang2010, Taniguchi2013, Tan2020, Liu2021}.
We expect that deep learning-based phase mapping can be applied to various forms of data.
However, it should be noted that each application may require its own specialized training dataset and specific deep learning algorithm, despite the ability of these algorithms to generalize.
The routine use of the technique across many different laboratories will likely only be achieved with much larger and more diverse training datasets (including various species and experimental conditions). 
Further, neural networks are not a filtering technique per se, and will only be able to perform a particular task (e.g. denoising) if they are trained on adequate data.
It will be crucial in future applications that training data includes all the different features, which are necessary for the neural network to learn all the different desired tasks.
While we provided a proof-of-concept, we also acknowledge that here is still potential for improving the phase mapping and direct PS prediction overall, especially for use with experimental data. 
We found that the direct PS prediction was more prone to challenging data than the prediction of phase maps, especially with optical mapping data or noisy and sparsified simulation data. 
We anticipate that better direct PS predictions or even higher phase mapping accuracies could be achieved with both more and better training data and more advanced neural network architectures. 
We aim to address these issues in future research.}

\section{Conclusions}
We demonstrated that convolutional neural networks can be used to predict phase maps and rotor core positions or phase singularities (PS) of reentrant cardiac excitation wave dynamics in both voltage-sensitive optical maps of ventricular fibrillation and simulated data mimicking low-resolution and/or sparse multi-electrode mapping data. 
The predictions can be made almost instantaneously, robustly and with accuracies of about $95\%$, and can be performed even in the presence of strong noise and highly sparse or incomplete data. 
{Neural networks used for phase mapping of cardiac excitation waves are able to analyze data obtained in one species, even if they were trained on a different species, and can predict phase maps and PS with experimental data, even if they were trained solely with simlated data of electrical spiral wave chaos.}
In the future, our approach could be used in electro-anatomic mapping applications for the diagnosis of atrial fibrillation.

\section{Supplementary Material}
The Supplementary Videos can be found online: \href{https://gitlab.com/janlebert/phasenet-supplementary-materials-arxiv}{gitlab.com/janlebert/phasenet-supplementary-materials-arxiv}.
$\\$
\textbf{Supplementary Video 1:} Neural network predictions of phase maps from voltage-sensitive optical mapping video data. The recording shows action potential spiral vortex waves during ventricular fibrillation on the left ventricular surface of a porcine heart, see also Fig.~\ref{fig:FigureExperiments}.
The pixel-wise normalized transmembrane voltage is shown on the left (yellow: depolarized, blue: repolarized tissue). Center: smoothed ground truth phase map, which was obtained from the noisy optical maps using the Hilbert transform, see Fig.~\ref{fig:FigurePreprocessing}A) and section~\ref{sec:methods:trainingdata}. Right: phase map predicted by the neural network.
$\\$
{\textbf{Supplementary Video 2:} Comparison of the neural network input images for the prediction of a single phase map for the pig, rabbit, and simulation datasets used in Figs.~\ref{fig:FigureSimulationToExperiment} and \ref{fig:FigureCrossTraining}.
The video shows the $N_t=10$ images given as input to the neural network to predict a single phase map for each type of dataset.}
$\\$
{\textbf{Supplementary Video 3:} PS prediction for rabbit optical mapping data using neural network models M1 (left), M1A  (middle), M1B (right). The PS for model M1 are predicted indirectly by first predicting phase maps then computing PS in the phase maps using the circular line integration method.}
$\\$
\textbf{Supplementary Video 4:} Phase maps predicted by the neural network from (sparse) simulated electrical excitation wave maps without and with noise ($\sigma = 0.3$). Left: electric excitation wave maps uses as network input ($N_t = 5$). Center: ground truth or true phase. Right: neural network output.
$\\$
\textbf{Supplementary Video 5:} Phase singularities (PS) predicted by the neural network M1A from (sparse) simulated electrical spiral wave chaos without and with noise ($\sigma = 0.2$) for $N_t = 5$. Left: ground truth electrical excitation, Center: network input, Right: predicted PS (black) and true PS (white) superimposed onto the corresponding electrical excitation wave maps.
$\\$
\textbf{Supplementary Video 6:} Phase maps predicted by the neural network from (sparse) simulated electrical excitation wave maps without and with noise ($\sigma = 0.3$) for $N_t = 5$.
$\\$
\textbf{Supplementary Video 7:} Neural network prediction of a single phase map. The 5 noisy and sparse electrical wave frames given as input to the neural network, as well as the predicted phase map and the true phase map are shown.
$\\$
{\textbf{Supplementary Video 8:}  PS prediction using model M1 for the sparse and noisy excitation patterns shown in Fig.~\ref{fig:FigureResults3Phase}. The predicted PS are shown in red, the ground truth in white. Left: The sparse excitation wave pattern with noise used as neural network input. Right: Ground truth excitation with sparsification shown in black.}

\section{Data Availability Statement}
The data that support the findings of this study are available from the corresponding author upon reasonable request.

\section{Funding}
This research was funded by the University of California, San Francisco. NR is a research Fellow supported by the Sarnoff Cardiovascular Research Foundation.

\section{Author Contributions}
JL and JC conceived the research and implemented the algorithms. JL, NR and JC conducted the data analysis and designed the figures. FHF provided the experimental data. JL, FHF and JC discussed the results and wrote the manuscript.

\section{Conflict of Interest}
The authors declare that the research was conducted in the absence of any commercial or financial relationships that could be construed as a potential conflict of interest.

\bibliographystyle{frontiersinHLTH&FPHY}
\bibliography{references}

\end{document}